\documentclass[11pt]{article} 
\usepackage{latexsym}
\usepackage{a4}
\usepackage[dvips]{graphicx}
\usepackage{psfrag}
\usepackage{bbm}      
\newcommand{\no}{\noindent}
\newcommand{\be}{\begin{eqnarray}}
\newcommand{\ee}{\end{eqnarray}}

\begin{document} 
\begin{titlepage}
\mbox{ UNITU-THEP-17/2003}
  \vspace{1.5cm}


\renewcommand{\thefootnote}{\fnsymbol{footnote}} 

\begin{center}
\Large\bfseries
Center Vortex Model for the Infrared Sector of $SU(3)$ Yang-Mills Theory --
Confinement and Deconfinement
\end{center}
\vspace{1cm}


\renewcommand{\thefootnote}{\fnsymbol{footnote}} 
\centerline{ M.~Engelhardt, 
             M.~Quandt\footnote[1]{Supported by 
             \emph{Deutsche Forschungsgemeinschaft} under contract 
             \# DFG-Re 856/4-2.}, 
             H.~Reinhardt${}^\ast$\vspace{1.5cm}}
\centerline{\textit{Institut f\"ur Theoretische Physik, 
             Universit\"at T\"ubingen} }
\centerline{\textit{D--72076 T\"ubingen, Germany.}} 
\renewcommand{\thefootnote}{arabic{footnote}} \vspace{1.5cm}


\begin{abstract}
The center vortex model for the infrared sector of Yang-Mills theory, previously
studied for the $SU(2)$ gauge group, is extended to $SU(3)$. This model is based
on the assumption that vortex world-surfaces can be viewed as random surfaces in
Euclidean space-time. The confining properties are investigated, with a
particular emphasis on the finite-temperature deconfining phase transition.
The model predicts a very weak first order transition, in agreement with
$SU(3)$ lattice Yang-Mills theory, and also reproduces a consistent behavior
of the spatial string tension in the deconfined phase. The geometrical
structure of the center vortices is studied, including vortex branchings, which
are a new property of the $SU(3)$ case.
\end{abstract}


\end{titlepage}


\section{Introduction} 
\label{sec:1}
\no
The vortex picture of the Yang-Mills vacuum was initially proposed
\cite{R9,R10,R11a,R11b} as a possible mechanism of confinement.
It is based on the idea that the presence of vortex flux randomly
distributed in space-time is sufficient to disorder Wilson loops to such
an extent that an area law behavior of their expectation values is generated.
The picture was further elaborated by the Copenhagen group \cite{R7}, who
considered the energetics of vortex formation in the Yang-Mills vacuum,
and it was also cast in lattice gauge theory terms \cite{R10}.
In particular, an alternative confinement criterion based on vortex free
energies on the lattice was formulated.

Apart from further developments within the lattice description
\cite{tomboulis}, the vortex picture lay dormant until the advent of new
gauge fixing techniques which permit the detection of center vortex structures
in lattice Yang-Mills configurations \cite{R8,alexandrou}. In essence,
one uses the gauge freedom to cast a given gauge configuration as accurately as
possible in terms of an appropriate center vortex configuration \cite{R2}. In
a second step, center projection, one extracts the vortex content by discarding
residual deviations from the aforementioned center vortex 
configuration \cite{R8}. Based on this procedure, it was possible to isolate 
the effects of the resulting \emph{center projection vortices} on strong 
interaction phenomena. It was soon realized that these center projection
vortices are physical in the sense that their density shows the proper 
scaling behaviour \cite{R8,X}.
Detailed exploration sharpened the understanding of 
the center gauge fixing techniques \cite{R2}, also indicating points where 
care must be taken in their application \cite{drama}.
Nevertheless, a broad body of evidence now indicates that the infrared 
properties of Yang-Mills theory can be accounted for in terms of vortex
effects; this includes not only the confinement properties (i.e., the
original focus of the vortex picture) \cite{R8,X,R14,jeffrev,cdom3,R1,pepe,X3},
but also the topological properties \cite{R2,R3,X2,fbme,R4} determining the
axial flavor $U_A (1)$ anomaly \cite{R13,bertle,R4} and the spontaneous
breaking of chiral symmetry \cite{R13,alexandrou,R5}.

These findings suggest that center vortices are the relevant infrared degrees
of freedom of Yang-Mills theory. Using that as the basic assumption, a
random vortex world-surface model for the infrared sector of Yang-Mills
theory was introduced \cite{R1} to complement the lattice gauge
investigations highlighted above. Initially studied for the $SU(2)$ gauge
group, it not only reproduces the confinement properties \cite{R1} of $SU(2)$
Yang-Mills theory quantitatively, but also the
topological susceptibility \cite{R4} and the spontaneous breaking of chiral
symmetry \cite{R5}.

The present work extends the model to the $SU(3)$ gauge group, focusing on
the confinement properties. The main physical difference to the $SU(2)$
case is the fact that vortex world-surfaces can branch due to the
existence of two types of quantized vortex flux instead of one. This is
expected to lead to important phenomenological consequences, such as
a change in the order of the finite temperature deconfinement phase
transition.

The paper is organized as follows: In section \ref{sec:2}, the center vortex
model of ref. \cite{R1} is revisited and extended to the gauge group $SU(3)$.
In section \ref{coupling}, the confinement properties in the parameter space of
coupling constants are investigated, yielding a determination of a physical
line in that space. The order of the deconfinement phase transition is studied
in section \ref{phase} and the geometrical structure of the world-surfaces
present in the vortex ensemble is investigated in section \ref{cluster}.
Section \ref{split} is devoted to the study of vortex branchings. Section
\ref{conclusions} contains concluding remarks.


\section{Definition of the random vortex world-surface model}
\label{sec:2}

\subsection{General properties of center vortices}
\label{sec:2.1}

Center vortices are closed lines of chromomagnetic flux in three space
dimensions; correspondingly, they are described by two-dimensional
closed world-surfaces in four-dimensional space-time. Their flux is
quantized according to the center of the underlying  gauge
group $SU(N)$. To be specific, this means that measuring the flux by
evaluating a Wilson loop $\mathcal{C} $ encircling the vortex\footnote{In
a less heuristic manner of speaking, one evaluates a Wilson loop which has
unit linking number with the (closed) vortex.} will result in a center
element of $SU(N)$,
\begin{equation}
W[\mathcal{C}] = \frac{1}{N} \mbox{ Tr } {\cal P} \exp 
\left( i\oint_{\mathcal{C}} A_{\mu } dx_{\mu } \right) \in Z(N) \ .
\label{wilsloop}
\end{equation}
In the case of the gauge group $SU(2)$ \cite{R1}, there is one nontrivial
center element and, correspondingly, one type of vortex flux; by contrast,
in the case of $SU(3)$, there are two nontrivial center elements in addition
to the trivial unit element,
\begin{equation}
Z(3) = \left\{ 1, \, \exp(2\pi i / 3) , \,
\exp(4\pi i / 3) \right\} \ .
\label{centel}
\end{equation}
This means that there are two distinct types of vortex flux. If one denotes
the center elements, eq.~(\ref{centel}), as
\begin{equation}
z_q = \exp (2\pi i q/ 3)
\end{equation}
with the {\em triality} $q$ defined modulo 3,
\begin{equation}
q \in \{ 0,1,2 \}
\end{equation}
then one can label the two types of vortex flux by associating them with
$q=1$ or $q=2$.

It is important to note that reversing the direction of the loop integration
in a Wilson loop or, equivalently, reversing the space-time orientation of
the vortex flux encircled by a given Wilson loop leads to a complex conjugation
of the latter. Since $z_1 = z_2^{\dagger } $, this means that a vortex flux
with $q=1$ is equivalent to an oppositely oriented vortex flux with $q=2$.
A general vortex world-surface configuration in fact contains $q=1$
surface patches and $q=2$ surface patches which share an edge at which the
space-time orientations of the patches are opposite to one another. 
The magnetic flux is discontinuous at such an edge in a way which may
be attributed to a Dirac monopole line on the vortex 
world-surface\footnote{This is quite analogous to the $SU(2)$ case, 
where monopoles form the boundary between vortex world-surface segments 
of opposite orientation \cite{R4}.}. Note that Dirac monopoles can be 
generated by pure gauge transformations in non-Abelian gauge theory
\cite{R2}; this must be taken into account when formulating the Bianchi
identity, cf.~eq.~(\ref{bianchi}) below.

However, for the purpose of evaluating the confinement properties of the vortex
ensemble, which are encoded in expectation values of Wilson loops, the
distinction between a $q=1$ vortex flux and an oppositely oriented $q=2$
vortex flux is irrelevant and the positions of monopoles on the
vortex world-surfaces thus do not have to be
considered\footnote{Correspondingly, in the following, the equivalent
labelings $q=-1$ and $q=2$ will both be used on an equal footing; 
after all, $z_{-1} = z_2 $.}. In particular, the action of the vortex
ensemble will be symmetric with respect to the two types of vortex flux.
For the purpose of evaluating the topological properties of the vortex
ensemble, on the other hand, the aforementioned distinction is indeed
important \cite{R2}, \cite{R3} (cf.~\cite{R4} for the $SU(2)$ model).
This topic is deferred to future work.

Focusing on the confinement properties, large Wilson loops may, of course,
be \emph{multiply} linked to vortices in any given vortex configuration. Due 
to the center quantization of the fluxes, however, such a large Wilson loop
can be decomposed into smaller ones in a straightforward fashion: 
Choose an arbitrary surface spanned by the Wilson loop, partition 
it into two subsurfaces and calculate the two Wilson loops which span the 
two subsurfaces\footnote{The line integrations must be oriented such that
they coincide with the orientation of the original Wilson loop and
such that integration paths common to the the two loops are integrated
over with opposite orientation, respectively.}. The product of these two loops
equals the original Wilson loop since, due to the center quantization of
the flux, any vortex configuration can be described by a
gauge field defined in the (Abelian) Cartan subalgebra of the gauge group 
\cite{R2}. This partitioning can be iterated until each elementary Wilson 
loop circumscribes at most one vortex flux and thus can be evaluated 
directly as indicated by (\ref{wilsloop}); the value of the large 
Wilson loop is found by adding the trialities from the elementary vortex 
fluxes.

Such a procedure can be realized in a very straightforward manner 
in the hypercubic lattice setting which will be employed throughout this
work. Given a Wilson loop on a hypercubic lattice and a surface spanned by 
the loop, this surface can be subdivided into elementary squares 
(plaquettes) on the lattice. Each of these plaquettes is pierced by at 
most one vortex; it should be stressed that the notion of ``piercing'' or 
``linking'' requires the lattices on which Wilson loops are defined 
and the lattices on which vortex world-surfaces are defined 
(composed of elementary squares) to be dual\footnote{This means that they have 
the same lattice spacing $a$, but are displaced from one another by the vector 
$(a/2,a/2,a/2,a/2)$.}
to one another. The precise triality labeling of the elementary squares
making up vortices and the consequent effect on a plaquette which is part
of a Wilson loop tiling will be specified further below.

\begin{figure}[t]
\centerline{
\hfill
\begin{minipage}{5cm}
\includegraphics[width = 5cm]{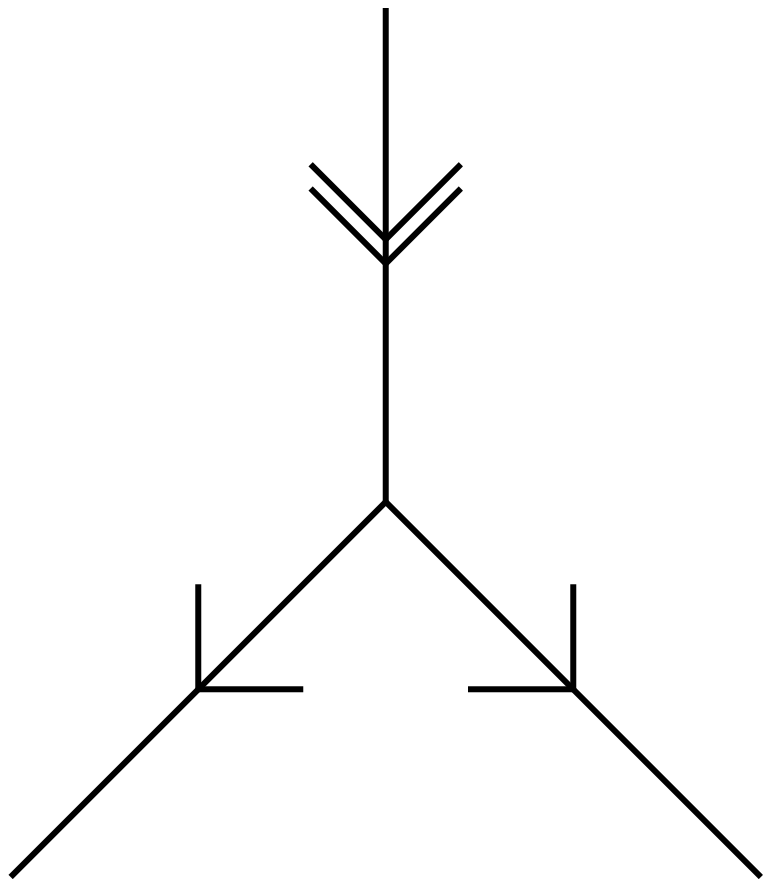} \end{minipage}\hfill 
\begin{minipage}{6cm}
\includegraphics[width = 6cm]{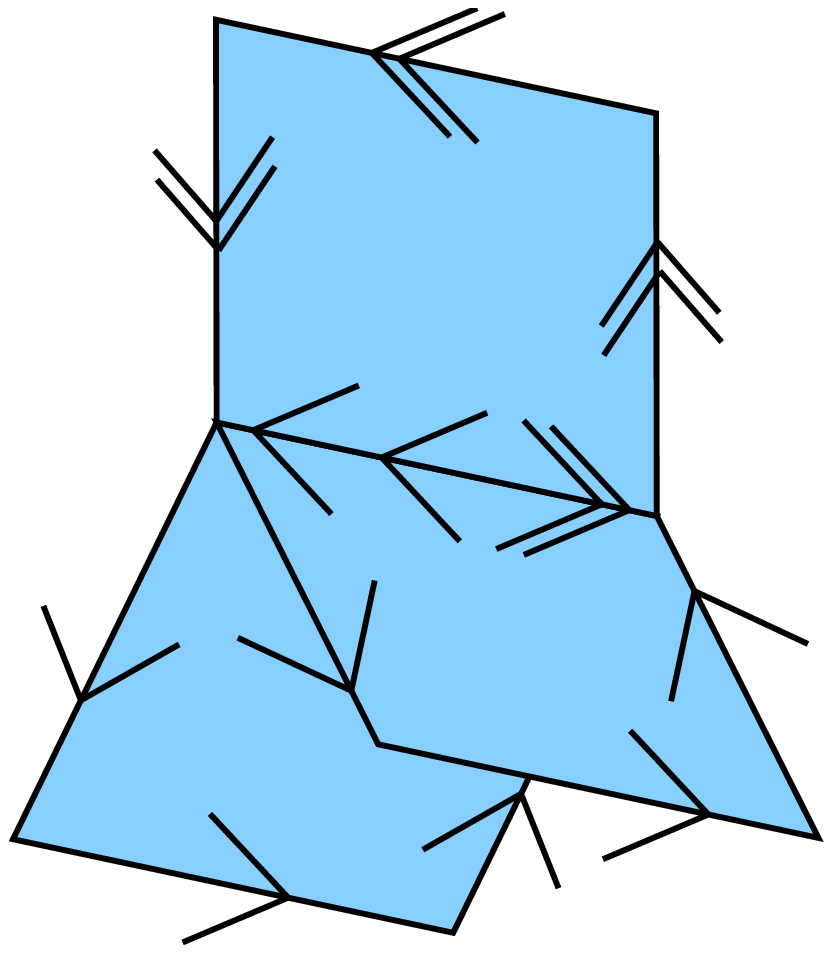}\end{minipage}
\hfill
}
\caption{Branching of $SU(3)$ center vortices: The left-hand panel depicts a
three-dimensional subspace, where a single vortex line with triality 
$q=2$ (double arrow) branches into two vortices of triality $q=1$ each
(single arrows). The right-hand panel displays a four-dimensional hypercubic
lattice view, where vortex surfaces, consisting of elementary squares,
branch along links. Elementary squares are associated with curls (see
main text), which must cancel on each link (modulo 3) according to
the Bianchi identity.}
\label{branch_fig}
\end{figure}

To complete the characterization of center vortices, it must be taken
into account that they are constrained by Bianchi's identity
\begin{equation}
\epsilon_{\mu \nu \kappa \lambda } \partial_{\nu }
F_{\kappa \lambda } \ = \ 0 \ \,
\mbox{modulo Dirac monopoles}
\label{bianchi}
\end{equation}
where it has already been used that vortex gauge fields can be described
completely within the Cartan subalgebra of the gauge group, as mentioned above.
Bianchi's identity expresses the fact that the (vortex) flux must be 
continuous, up to the Dirac monopole sources and sinks already mentioned 
earlier (which are allowed due to the compact nature of the gauge group).
It is possible to construct a field strength corresponding to an 
arbitrary vortex world-surface configuration \cite{R2} and discuss the 
consequences of eq.~(\ref{bianchi}) for vortex world-surface topology;
here, however, a more inductive description shall suffice. 
In three-dimensional space, the consequences of continuity of flux are 
intuitively clear: Vortex lines are closed, since the flux would be
discontinuous at an open end. This is analogous to the $SU(2)$ case. 
In contrast to $SU(2)$, however, the generalisation to $SU(3)$ offers the 
additional possibility of a $q=2$ vortex flux branching into two $q=1$ vortex 
fluxes, cf.~fig.~\ref{branch_fig}. In the following, this phenomenon
will be referred to as \emph{vortex branching}.

In terms of the vortex world-surfaces in four-dimensional space-time
(thought of as composed of elementary squares on a hypercubic lattice),
this translates into a condition on each lattice link, 
cf.~fig.~\ref{branch_fig}. Note that the left-hand panel of
fig.~\ref{branch_fig} corresponds to the right-hand panel taken e.g.~at
a fixed lattice time.

In the four-dimensional picture, one chooses
an orientation associated with each elementary square by defining
a sense of curl, i.e., a sense in which the links bordering the square
are run through, cf.~fig.~\ref{branch_fig}. Additionally, each of these curls
is endowed with a magnitude, namely the triality $q=1$ or $q=2$ describing the
vortex flux carried by the elementary square. Then the Bianchi identity
corresponds to a constraint to be satisfied on each lattice link:
The sum of the curls attached to the link (with opposite orientation
implying opposite sign of the contribution) must vanish modulo 3, as
happens on the central link in the right-hand panel of fig.~\ref{branch_fig}.
It is not necessary at this point to undertake any effort to cast
this condition in formal language; as will become clear below, in 
practice there will be no need to check the Bianchi constraint for
any vortex configurations since the Monte Carlo update procedure
generating the vortex world-surface ensemble will be such that the
Bianchi identity is guaranteed at every step.

\subsection{Center vortex model dynamics} 
\label{prop}

For the case of the gauge group $SU(2)$, an effective model for infrared
Yang-Mills dynamics based on center vortex degrees of freedom was
presented in \cite{R1}, where the physical motivation for such random
vortex world-surface models is discussed in depth. In addition to the
confinement properties studied in \cite{R1}, the model also turns out
to successfully describe the topological properties of $SU(2)$ Yang-Mills
theory \cite{R4} and the spontaneous breaking of chiral symmetry
\cite{R5}. The physical principles underlying this vortex model can also
be applied to the gauge group $SU(3)$. Adapted to the $SU(3)$ case,
they are:

\begin{enumerate}
\item \underline{\textsl{Description of vortex world-surfaces:}}
Vortex world-surfaces will be described by composing them of elementary
squares on a hypercubic lattice, as already mentioned above. Each
elementary square on the hypercubic lattice will be associated with a
triality $q_{\mu \nu } (x) \in \{-1,0,1\} $, where $x$ denotes the
lattice site from which the elementary square extends into the positive
$\mu $ and $\nu $ directions. The ordering of the indices corresponds
to the curl already mentioned in the previous section in connection
with the Bianchi identity: The value of $q_{\mu \nu } (x)$ specifies
the value of the curl corresponding to starting out from $x$ into the
positive $\mu $ direction, and then onwards around the elementary
square. For definiteness, the indices will usually be ordered such that
$\mu < \nu $. If for notational convenience $q_{\nu \mu } (x)$ with
$\nu > \mu $ is quoted instead, then this is defined as
$q_{\nu \mu } (x) = -q_{\mu \nu } (x)$ in accordance with the remarks
in the previous section. The value $q_{\mu \nu } (x) =0$ means that the
elementary square is not part of a vortex surface, whereas the values
$q_{\mu \nu } (x) =\pm 1$ label the two possible types of vortex
flux which may be carried by the elementary square.

Having thus specified the labeling of the elementary squares making
up the vortex surfaces, one can now also give their precise effect on
the elementary Wilson loops (plaquettes) $U_{\kappa\lambda}(y)$ they
pierce. It should be emphasised once again that Wilson loops are defined
on a lattice \emph{dual} to the one on which the vortex surfaces are
constructed. In analogy to the description of vortex elementary squares
given above, the elementary Wilson loop $U_{\kappa\lambda}(y)$ thus refers
to a plaquette extending
from $y$ into the positive $\kappa $ and $\lambda $ directions, with the
integration oriented such that one starts at $y$, integrates first
into the positive $\kappa $ direction, and then onwards around the
plaquette. The plaquette $U_{\kappa \lambda } (y)$ is pierced precisely
by the dual lattice elementary square $q_{\mu \nu }(x)$, where the
indices $\kappa , \lambda , \mu , \nu $ span all four space-time
dimensions and $x=y+(\vec{e}_{\kappa } + \vec{e}_{\lambda } -
\vec{e}_{\mu } - \vec{e}_{\nu } ) a/2$, with $a$ denoting the lattice
spacing. $U_{\kappa \lambda } (y)$ can thus be given exclusively in
terms of the corresponding $q_{\mu \nu } (x)$, namely
\begin{equation}
U_{\kappa \lambda } (y) = \exp \left( i \pi / 3 \cdot
\epsilon_{\kappa \lambda \mu \nu } \,q_{\mu \nu }(x)\right)
\end{equation}
(with the usual Euclidean summation convention over Greek indices).
\item \underline{\textsl{Transverse Thickness:}} In the full Yang-Mills 
theory, vortices possess a physical thickness perpendicular to the 
vortex surface, i.e.~they are non-singular field configurations with 
finite action \cite{R8,casimir}. They resemble Nielsen-Olesen
vortices \cite{R7}
rather than infinitely thin magnetic flux tubes. In particular, this means
that vortices cannot be packed arbitrarily densely: If two parallel vortices
approach each other closer than the average transverse thickness of their
profile, they become undistinguishable from a single vortex with the triality
of the two component vortices summed. In the $SU(2)$ model, this means that
they annihilate and become equivalent to the vacuum. In the $SU(3)$ case,
however, the $Z(3)$ triality algebra, eq.~(\ref{centel}), allows e.g.~for two
$q=1$ vortices to combine into a single $q=2$ vortex if they come
sufficiently close.

In the present vortex model, this behavior is reflected in a fixed, finite 
physical value of the lattice spacing $a$ which mimics transverse vortex
extension: Parallel vortices at a distance smaller than $a$ are  \emph{not}
resolved and instead are replaced by a single vortex of the summed triality.

This should be contrasted with the center projection vortices \cite{R8,R2}
which can be extracted from lattice gauge configurations on arbitrarily fine
lattices. They represent only a rough localisation of physical thick vortices
and fluctuate rapidly on all wavelengths even if the underlying thick
vortex structure is smooth. In addition, the exact position of a center
projection vortex within the thick profile of the physical vortex it is
extracted from depends on the details
of the gauge fixing procedure \cite{R2}. The center projection vortex
effective action therefore presumably is very complicated and in
particular non-local as the cutoff diverges into the ultraviolet,
while a true low energy effective theory should only retain degrees
of freedom which fluctuate on wavelengths up to a certain upper limit
(provided here by the lattice spacing). In this spirit, the vortices
described by the present model, though formally still composed of
infinitely thin elementary squares\footnote{One possible refinement of the
model presented here would be to introduce an explicit transverse smooth
field strength profile \cite{fbme} for vortices; this would be
necessary e.g.~to describe the medium-range Casimir scaling behavior
of adjoint representation Wilson loops \cite{casimir}.}, represent the smooth
centers of the profiles of physical (thick) vortices rather than the 
aforementioned rapidly fluctuating thin projection vortices.
Alternatively, one may think of the model vortices treated here
as thin center projection vortices with the short distance fluctuations
averaged out down to a scale $a$ set by the fixed lattice cutoff. This
point of view entails that the vortex action should be well described
by a few local terms in the spirit of a derivative expansion
\cite{R2}.

\item\underline{\textsl{Vortex Action:}} Vortex creation costs a certain
action per unit area. This is the first term expected from a derivative 
expansion \cite{R2},
\begin{equation}
S_{\rm area}[q] = \epsilon \sum_{x}\sum_{\mu,\nu \atop \mu<\nu}
\left| q_{\mu\nu}(x) \right| \, ,
\label{area}
\end{equation}
where this form of the action assumes that the triality labelings have
been chosen such that $q_{\mu \nu } (x) \in \{ -1,0,1 \} $. Note thus
that $S_{\rm area} $ is symmetric with respect to the two possible
types of vortex flux.

Vortices are also \emph{stiff}, 
i.e.~there is a penalty in the action for bends in the vortex surfaces.
To be specific, an action increment $c$ is incurred for each pair of 
elementary squares in the lattice which share a link and are part of a
vortex, but do not lie in the same plane. This can be expressed in terms
of a sum over links,
\begin{eqnarray}
S_{\rm curv}[q] \! \! \! \! &=& \! \! \! \!
c \sum_x\sum_\mu \Bigg[ \sum_{\nu < \lambda \atop \nu \neq \mu,
\lambda\neq \mu} \Big( | q_{\mu\nu}(x) \, q_{\mu\lambda}(x) |
 + | q_{\mu\nu}(x) \, q_{\mu\lambda}(x-e_\lambda) | \label{curvature} \\
& & \ \ \ \ \ \ \ \ \ \ \ \ \ \ \ \ \ \ 
+ | q_{\mu\nu}(x-e_\nu) \, q_{\mu\lambda}(x) |
 + | q_{\mu\nu}(x-e_\nu) \, q_{\mu\lambda}(x-e_\lambda) |
\Big)\Bigg] \nonumber \\
&=& \! \! \! \! \frac{c}{2} \sum_{x}\sum_\mu \left[ \left[
\sum_{\nu\neq\mu} \left( | q_{\mu\nu}(x) | + 
| q_{\mu\nu}(x-e_\nu) | \right) \right]^2 \! \! \! - \! \!
\sum_{\nu\neq\mu} \! \! \Big[ | q_{\mu\nu}(x) | + 
| q_{\mu\nu}(x-e_\nu) | \Big]^2 \ \right] \ . \nonumber
\end{eqnarray}
As in the case of $S_{\rm area} $, these expressions for
$S_{\rm curv} $ assume that the triality labelings have
been chosen such that $q_{\mu \nu } (x) \in \{ -1,0,1 \} $. Also
$S_{\rm curv} $ is symmetric with respect to the two possible
types of vortex flux.

This is a plausible model assumption in view of the fact, already discussed
further above, that the two types of vortex flux are related by a space-time
inversion, at least as far as their effect on Wilson loops is concerned.
Nevertheless, it should be noted that the curvature action presented
above is not the most general one which involves elementary squares
sharing a link and which respects the aforementioned symmetry. The form
(\ref{curvature}) manifestly is a sum of terms depending only on pairs of
elementary squares attached to a given link; there are no higher order terms
simultaneously involving more than two elementary squares such as
$|q_{\mu \nu } (x) q_{\mu \lambda } (x) q_{\mu \kappa } (x)|$, which
would be entirely admissible\footnote{Such terms could e.g.~be used to
change the weighting of vortex branchings. Without them, the distribution
of vortex branchings is determined indirectly by the vortex dynamics and
the entropy of branched random surfaces. Vortex branchings will be discussed
in more detail in section \ref{split}.}. This type of truncation was
already assumed in the $SU(2)$ case \cite{R1}; it does not constitute
a new feature of the present investigation and serves to keep the number
of independent coupling constants small and the model predictive.

\item\underline{\textsl{Monte Carlo update procedure and continuity of flux:}}
The model vortex dynamics will be realized by generating an ensemble of random
vortex world-surfaces using a Monte Carlo procedure weighted with the action
presented above. The elementary update will be chosen such that continuity
of flux is guaranteed at every step. Namely, choose an elementary
three-dimensional cube in the lattice extending from a lattice site $x$
into the positive $\mu $, $\nu $ and $\lambda $ directions\footnote{In
practice, sweeps through the lattice are performed, considering updates
associated with each elementary cube in the lattice in turn.}. Update
the configuration by adding the flux corresponding to a vortex
shaped as the elementary cube surface to the flux previously present.
All six elementary squares making up the surface of the cube are thus
updated simultaneously as follows,
\begin{equation}
\begin{array}{l@{\qquad}l}
q_{\mu\nu}(x) \to q_{\mu\nu}(x) + w\,, &
q_{\mu\nu}(x+e_\lambda) \to q_{\mu\nu}(x + e_\lambda) - w \\
q_{\nu \lambda}(x) \to q_{\nu \lambda}(x) + w\,, &
q_{\nu \lambda}(x+e_\mu) \to q_{\nu \lambda}(x+e_\mu) - w \\
q_{\lambda\mu}(x) \to q_{\lambda\mu}(x) + w\,, &
q_{\lambda\mu}(x+e_\nu) \to q_{\lambda\mu}(x+e_\nu) - w
\end{array}
\label{update}
\end{equation}
with $w=\pm 1$ . In practice, the value $w$ with which the update is attempted
(i.e.~the triality of the superimposed cubic vortex) is chosen at random with 
equal probability. Since the linear superposition of two configurations which 
satisfy continuity of flux leads again to a configuration which satisfies 
that constraint, the update algorithm preserves Bianchi's identity,
eq.~(\ref{bianchi}), at every step while allowing to generate every valid 
vortex world-surface configuration.
\end{enumerate}


\begin{figure}[t]
\centerline{
\begin{minipage}{8cm}
\psfrag{xx}{$\epsilon$}
\psfrag{yyy}{$c$}
\psfrag{zzzzzz}{$\sigma_{0} \,a^2$}
\includegraphics[width = 8cm]{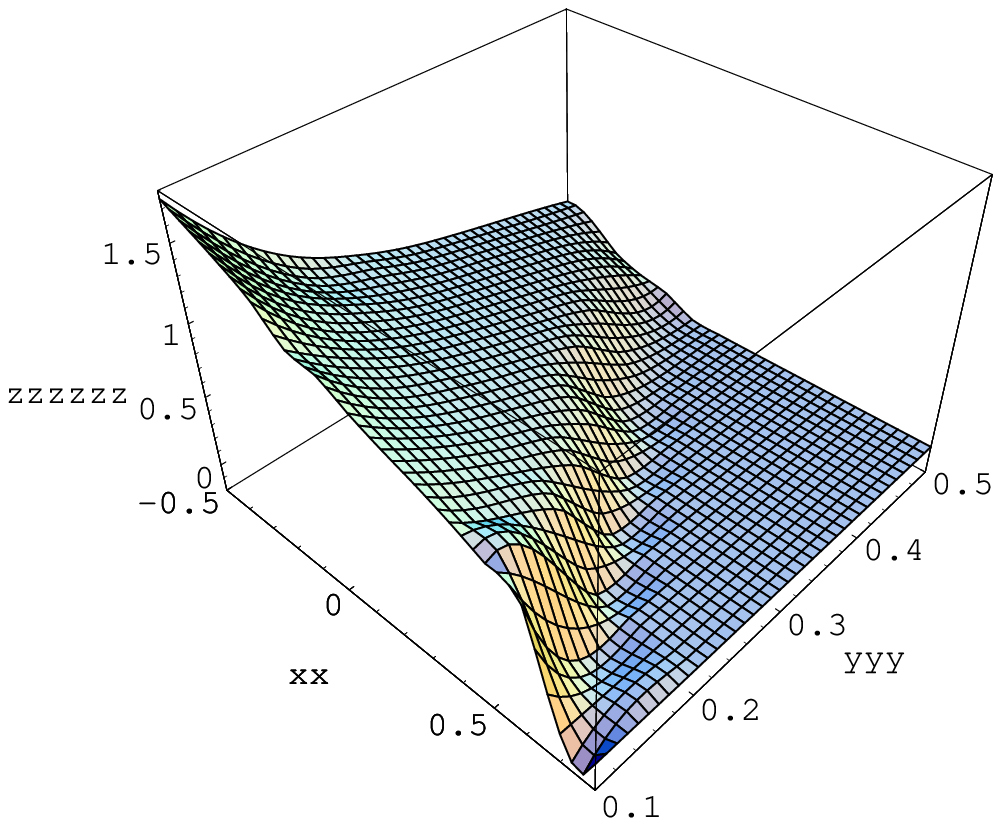} \end{minipage}\hfill 
\begin{minipage}{7cm}
\psfrag{eeeee}{$\epsilon$}
\psfrag{ccccccc}{$c$}
\includegraphics[width = 7cm]{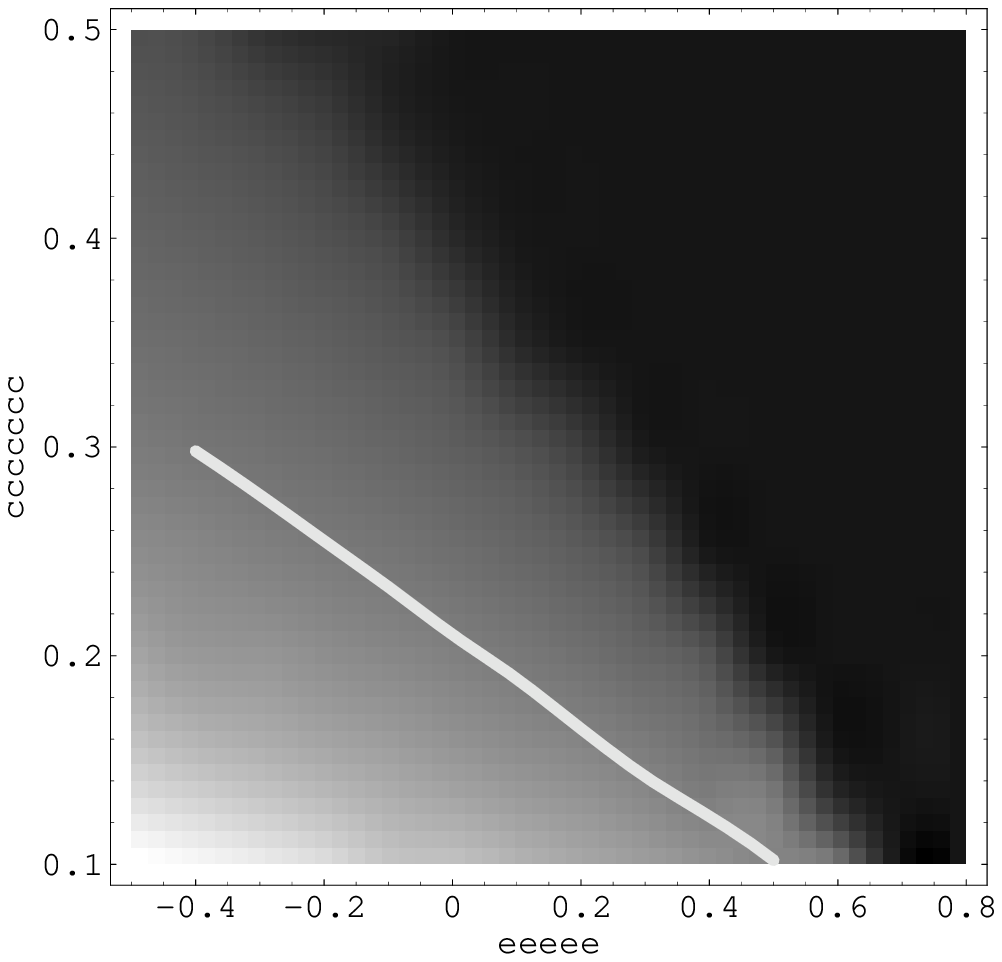}\end{minipage}
}
\caption{Plane of coupling constants $(\epsilon,c)$. The left-hand panel shows
the zero-temperature string tension $\sigma_{0} $ in units of the lattice
spacing, as a function of $\epsilon$ and $c$. In the right-hand panel, the
three-dimensional representation has been replaced
by a density plot in which brighter colours indicate higher string tension.
The thick white line in the confining region locates pairs $(\epsilon,c)$
at which the ratio $T_c/\sqrt{\sigma_{0} } \approx 0.63$ of full 
$SU(3)$ Yang-Mills theory is reproduced.}
\label{couplingspace}
\end{figure}

\section{Confinement properties and the space of coupling constants}
\label{coupling}
The model described in the previous section formally has three independent 
parameters, namely, the two dimensionless coupling constants $\epsilon$ and 
$c$ introduced in the action, and the dimensionful lattice spacing $a$.
As emphasised above, the spacing $a$ has a definite physical interpretation
and thus must be given a 
fixed physical value rather than being taken to zero as in conventional 
lattice gauge theories. Eventually, $a$ will be fixed by fitting the
zero-temperature string tension $\sigma_{0} $ to the phenomenological value
\begin{equation}
\sigma_{0} = (440\,\mathrm{MeV})^2\,.
\label{stringtension}
\end{equation}
With this arrangement, the lattice spacing $a$ is assigned a physical 
value which determines the transverse thickness of the vortices (see
previous section). Moreover, it provides the ultraviolet cutoff for the
largest momenta $|p| = \pi/a$ which can be resolved in this low energy
effective theory, and it sets the scale for all dimensionful quantities.

With the determination of $a$ understood, consider the phase
diagram of the model in the coupling constant ($\epsilon,c$) space.
Figure \ref{couplingspace} shows the result of Wilson loop measurements on a 
symmetric $16^4$ lattice, which represents the zero-temperature approximation.
From the Wilson loops, the string tension $\sigma_{0} $ can be extracted
e.g.~by computing \emph{Creutz ratios} or directly by fitting the area-law
fall-off. The result is a simple phase diagram containing a confining and
a non-confining region, cf.~the left-hand panel of fig.~\ref{couplingspace}.
The confining region is located at small $c$ and small or even negative
$\epsilon$, where the formation of a high density of connected, crumpled
vortex world-surfaces is favoured.

For a large range of coupling constants, one furthermore encounters a
deconfinement transition to a phase with vanishing string tension
when raising the temperature by decreasing the 
number of lattice spacings $N_0$ in the Euclidean time direction. 
At finite temperatures, the heavy quark potential must be extracted 
from Polyakov loop correlators. Since the model is Abelian, this is 
equivalent to measuring time-like $(N_0 \times R)$ Wilson loops with maximal 
extension in the Euclidean time direction. 

As emphasised above, the lattice spacing of the model is a 
fixed quantity that cannot be adjusted towards a continuum limit. As a 
consequence, changing $N_0$ while keeping $(\epsilon,c)$ fixed only
allows to alter the temperature in rather big steps. To obtain additional
data, an interpolation method \cite{R1} is used: For fixed $\epsilon$,
$N_0 = 1,2,3$ are considered in turn and $c$ is varied until the
deconfining phase transition is observed at a critical coupling
$c^\ast(N_0)$. This in turn yields the critical 
temperature $a T_c = 1/N_0$ for the pair $(\epsilon,c^\ast(N_0))$,
from which $a T_c(\epsilon,c)$ may be obtained for all couplings by 
interpolation.

With measurements, in lattice units, of the critical temperature, $aT_c$,
and the zero-temperature string tension, $\sigma_{0} a^2$, the physical ratio
$T_c /\sqrt{\sigma_{0} } $ can be computed
for all values of the coupling constants $\epsilon, c$. Measurements in full
$SU(3)$ lattice Yang-Mills theory favour the value $T_c/\sqrt{\sigma_{0} }
\approx 0.63$ for the gauge group $SU(3)$ \cite{SU3_Tc}. Taking this number as
a second input (besides the value eq.~(\ref{stringtension}) of the string 
tension) yields the white line in the right-hand panel of
fig.~\ref{couplingspace}. Note that points of the same colour in that plot
indicate the same string tension in lattice units $\sigma_{0} a^2 $,
i.e.~$\sigma_{0} a^2 $ does not vary considerably on the white physical line.
As $\sigma_{0} $ is being used to fix the lattice spacing, this entails
in turn that the spacing itself will be approximately constant along the 
physical line (it varies only by about 10 \%).

On the one hand, this corroborates the physical picture of the lattice
spacing as a fixed quantity determining the spatial resolution and the 
transverse thickness of the vortices. Further measurements, such as of the 
spatial string tension, provide additional evidence that physics is 
approximately constant along the white line in fig.~{\ref{couplingspace}}.
This means that one cannot use further physical input to find the unique
point in the phase diagram where the model matches low-energy Yang-Mills
theory: There is an entire line of such points which, to a good approximation,
are equivalent. This is discussed in more detail in \cite{R1} for the
$SU(2)$ case, where the same phenomenon occurs. For convenience, one specific
point on the physical trajectory will be chosen in the following and all
measurements will be performed with the parameters
\begin{equation}
\epsilon = 0\,,\qquad\qquad c = 0.21
\label{physical}
\end{equation}  
unless stated otherwise. For this choice of parameters, the lattice spacing
as extracted from eq.~(\ref{stringtension}) is
\begin{equation}
a = 0.39 \,\mathrm{fm}\, ,
\end{equation}
which equals the spacing found in the $SU(2)$ model \cite{R1}.
For other points on the physical trajectory, the lattice spacing deviates less
than 10 \% from this value, with lower $\epsilon$ giving slightly smaller $a$.

\begin{figure}[t]
\centerline{
\begin{minipage}{9cm}
\includegraphics[width = 9cm]{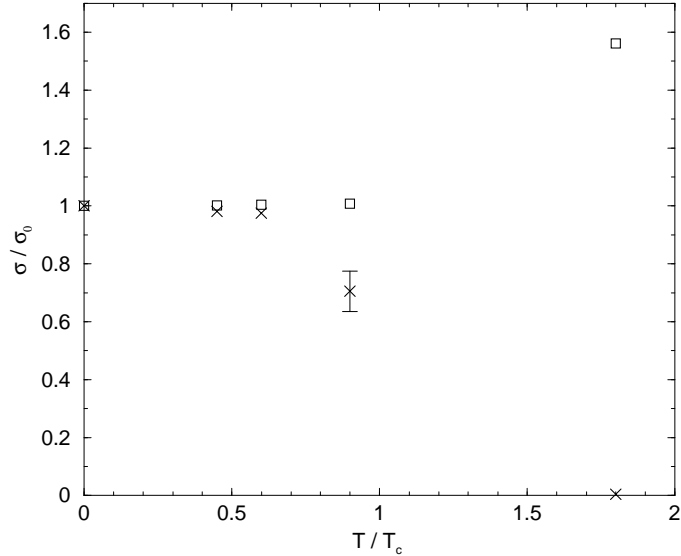} \end{minipage}
}
\caption{String tension between two static colour charges
(\emph{crosses}) and spatial string tension (\emph{squares}) as a
function of temperature. Measurements were taken on a $16^3 \times N_0 $
lattice for the physical choice of parameters, eq.~(\ref{physical}). The
error bar for the temporal string tension at $T/T_c = 0.90$ reflects a
systematic uncertainty resulting from subleading logarithmic corrections
in the potential near the phase transition \cite{karschlog}.}
\label{figure_spatial}
\end{figure}

Having chosen a physical set of parameters, one can begin to predict further
quantities. Fig.~\ref{figure_spatial} displays measurements of the string
tension between static colour sources as well as the spatial string tension
as a function of temperature. While the deconfinement temperature reflected
in the string tension between static colour sources has been used in fixing
the coupling constants, the behavior of the spatial string tension,
extracted from spatial Wilson loops, does not enter the construction of the
model and is thus predicted. Particularly the behavior at high
temperatures, where a finite spatial string tension persists, is interesting.
In the $SU(2)$ case \cite{R1}, the obtained model values
agreed with the ones measured in $SU(2)$ lattice Yang-Mills theory
to within 1\%; such high agreement certainly is coincidental in view of the
fact that those measurements take place near the ultraviolet limit of
validity of the vortex model. In the $SU(3)$ case,
cf.~fig.~\ref{figure_spatial}, the agreement at the highest temperature,
$T=1.80\,T_c $, is still impressive, namely to within 5\% compared with the
full $SU(3)$ Yang-Mills value obtained in \cite{SU3_Tc}. This is 
furthermore well within the error bars quoted in \cite{SU3_Tc}.

\begin{figure}[t]
\centerline{
\begin{minipage}{7.5cm}
\includegraphics[width = 7.5cm]{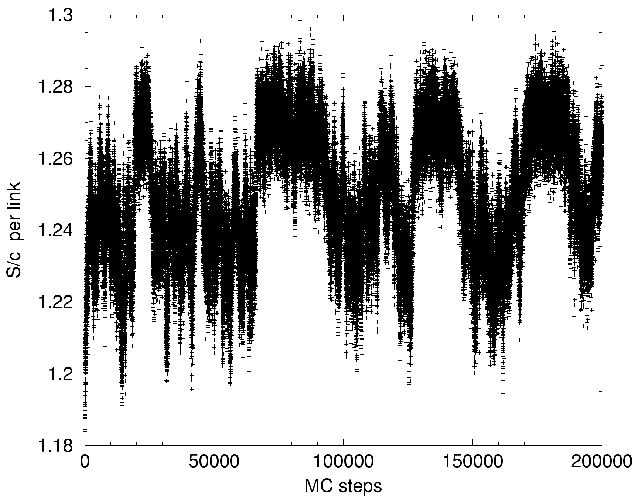} \end{minipage}\hfill 
\begin{minipage}{7.5cm}
\includegraphics[width = 7.5cm]{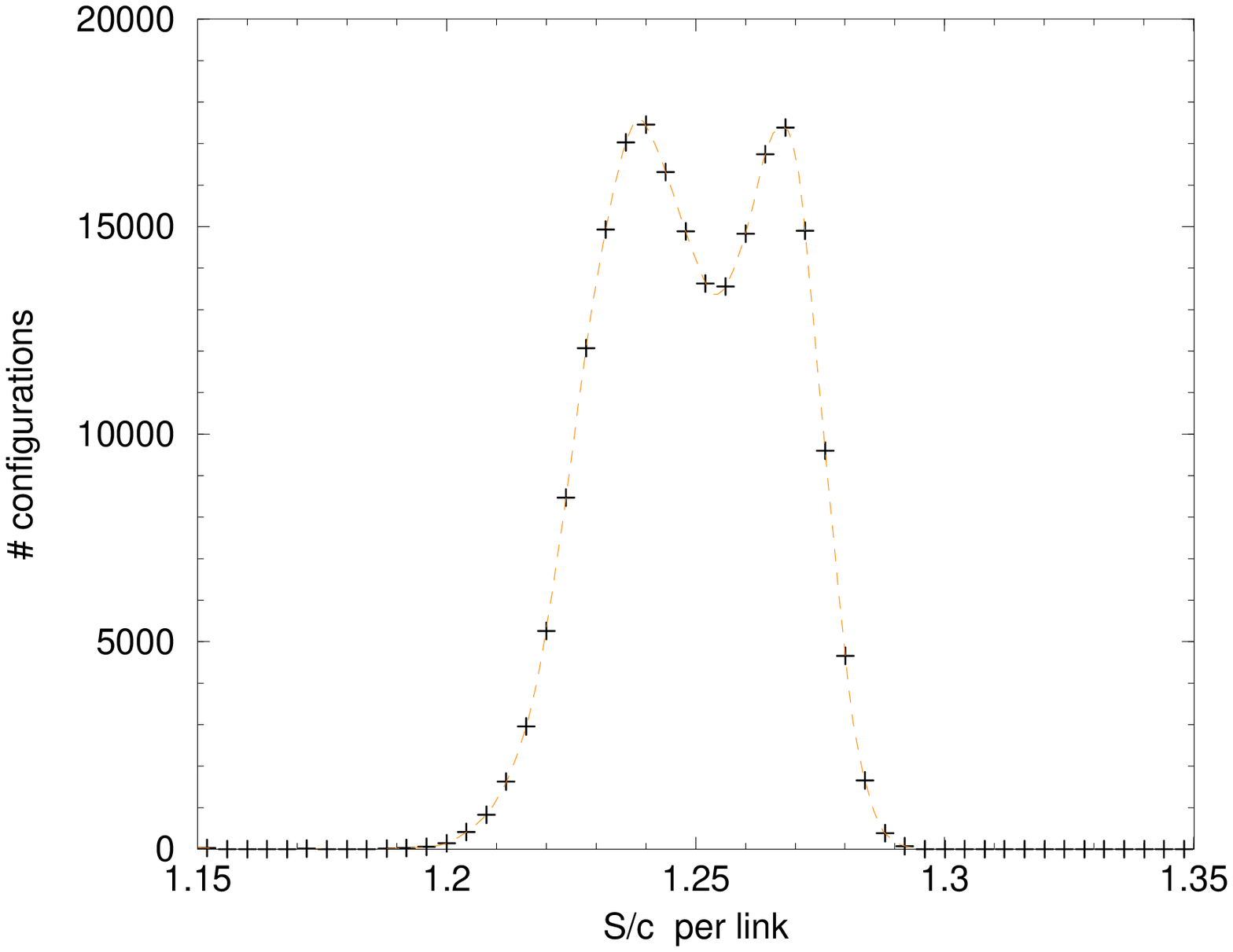}\end{minipage}
}
\caption{Left-hand panel: Action per link as a function of 
the Monte Carlo simulation time. Measurements were taken on a 
$30^3 \times 2$ lattice at the deconfinement phase transition point 
$\epsilon = 0$ and $c = 0.2359$. This set of parameters is close to
the physical trajectory, cf.~main text. The right-hand panel shows the
same data as a histogram of configurations binned according to their 
curvature action. The resulting action distribution exhibits the double
peak structure characteristic for a first order phase transition.
}
\label{fig2359}
\end{figure}

\section{Finite temperature phase transition}
\label{phase}
\begin{figure}[t]
\centerline{
\begin{minipage}{7.5cm}
\includegraphics[width = 7.5cm]{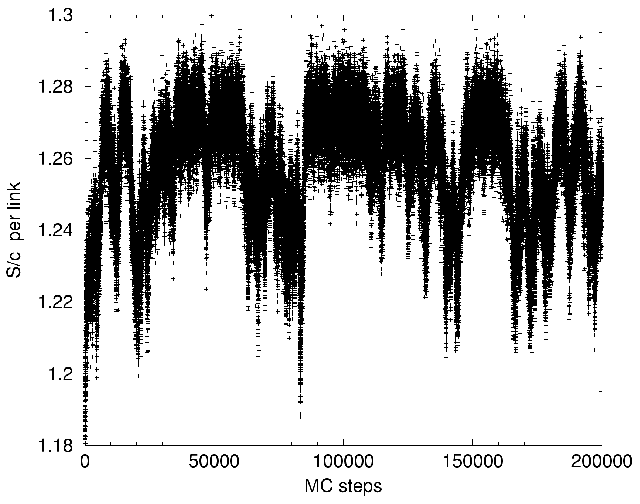} \end{minipage}\hfill 
\begin{minipage}{7.5cm}
\includegraphics[width = 7.5cm]{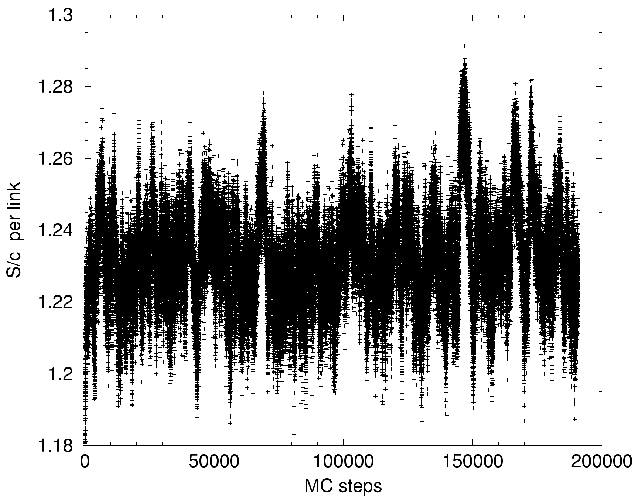}\end{minipage}
}
\caption{Action per link as a function of Monte Carlo time.
Measurements were taken on a $30^3 \times 2$ lattice with $\epsilon = 0$.
The value of the curvature coupling $c$ was chosen such that the system is
biased slightly towards the confined phase (left-hand panel, $c=0.2357$)
and the deconfined phase (right-hand panel, $c=0.2361$), respectively.}
\label{fig2357}
\end{figure}

\begin{figure}[t]
\centerline{
\begin{minipage}{7.5cm}
\includegraphics[width = 7.5cm]{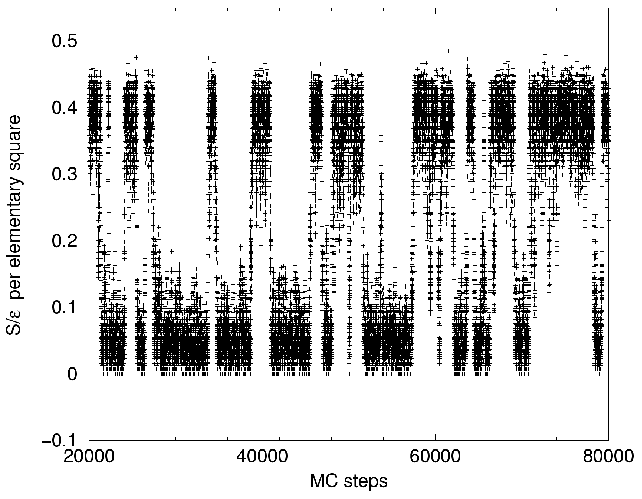} \end{minipage}\hfill 
\begin{minipage}{7.5cm}
\includegraphics[width = 7.5cm]{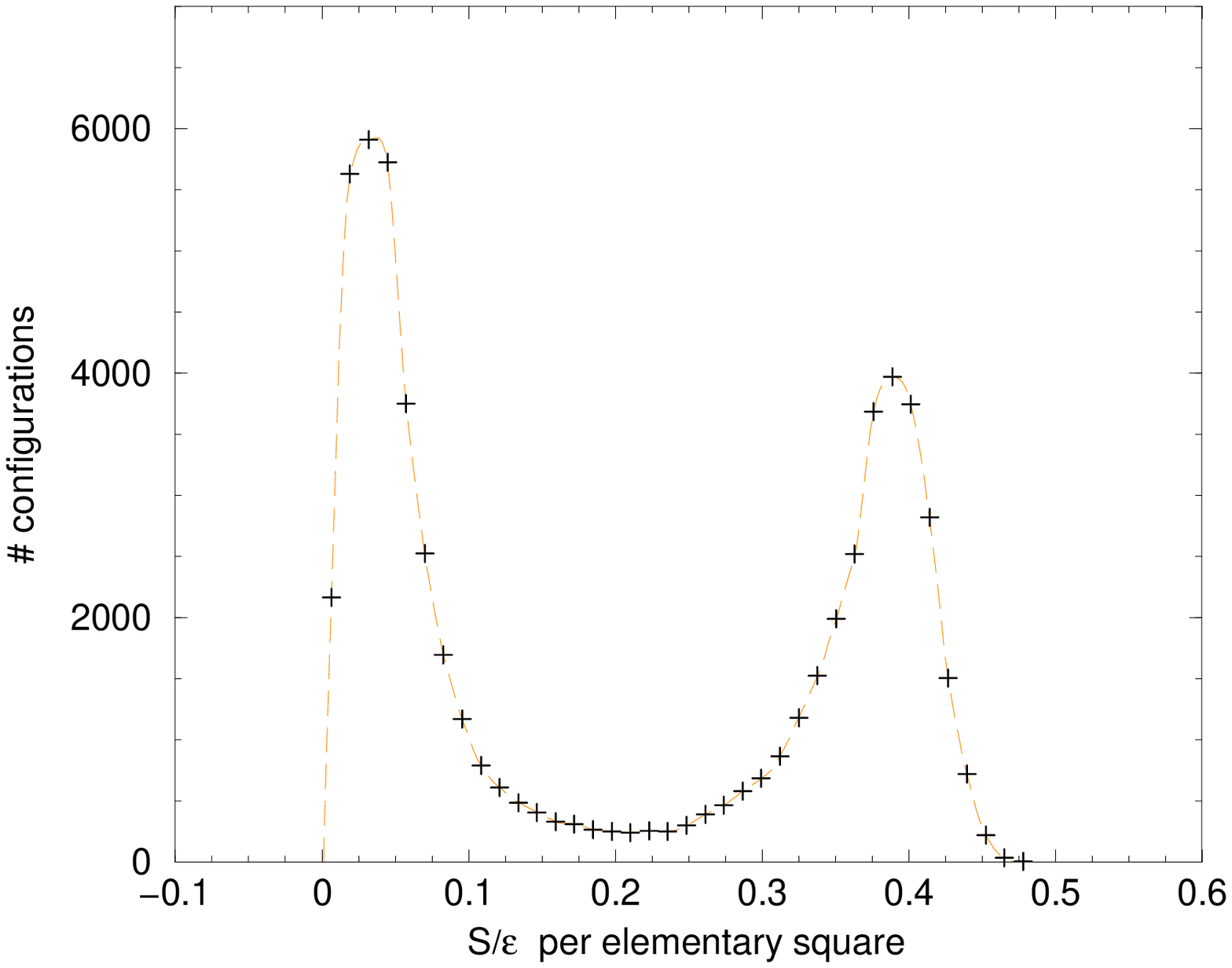}\end{minipage}
}
\caption{Left-hand panel: Action per elementary square as a function of Monte
Carlo time at the (unphysical) point $c=0$ and $\epsilon=0.993$. The
corresponding action distribution in the right-hand panel exhibits a clear
signal of a strong first order phase transition for this
unphysical choice of parameters, even on very small lattices
(measurements were taken on a $4^3 \times 2$ lattice).} 
\label{small_histogram}
\end{figure}

A topic of particular interest is the order of the deconfining phase
transition. Ideally, one would be interested in studying the physical
point $c=0.21$, $\epsilon = 0$ 
and directly varying the temperature via the temporal
extension of space-time. In practice, only discrete values of the 
temporal extension are accessible directly at $c=0.21$; for this reason,
the authors instead worked at a fixed number of temporal lattice spacings
$N_0 =2$ and varied $c$ around the phase transition point, which is
located at $c=0.2359$ (on $30^3 \times 2$ lattices). This is rather close
to the physical value $c=0.21$ and thus is expected to furnish a good
indication of physics there. Note that results directly on the physical
trajectory could be obtained using an interpolation procedure \cite{R1}
based on studying the phase transition at several $N_0$ and the
associated critical values $c^{*} (N_0 )$; however, such an interpolation
would indeed be dominated by the results at $N_0 =2$, $c=0.2359$ given below.
The above accurate estimate of the critical value, $c=0.2359$, was
determined by finding the maximal slope of the curvature
action per lattice link\footnote{For measurements taken at $\epsilon=0$, 
the curvature action is, of course, identical to the total action.
Note that the former is determined by considering pairs of vortex
elementary squares which share a link but do not lie in the same plane.
Thus, the curvature action can be locally attributed to the links.
Fig.~\ref{small_histogram}, by contrast, shows the total action at a point
in the plane of coupling constants with $c=0$, where it equals the surface
action and can therefore be attributed to \emph{elementary squares} rather
than \emph{links}.} as a function of $c$. Subsequently, the authors
carried out long Monte Carlo runs on large, $30^3 \times 2$ lattices,
recording the curvature action per lattice link for every configuration.
Such a Monte Carlo history is depicted in fig.~\ref{fig2359}. It displays
the behavior characteristic of a first order phase transition, as 
is to be expected for a model with $Z(3)$ symmetry \cite{svetitsky}.
As the Monte Carlo process progresses, the value of the action per
lattice link at times fluctuates around a lower mean value (deconfined
phase) and at other times around an upper mean value
(confined phase). At $c=0.2359$, the two cases occur with
approximately equal weight. At the neighboring value $c=0.2357$,
the effect of the phase transition is still easily discernible, whereas
at $c=0.2361$, the evidence is already rather tenuous, cf.~fig.~\ref{fig2357}.

It is important to note the rather weak first-order behavior. The 
difference between the aforementioned mean values in the two phases
is small and it was necessary to go to very large, $(30a)^3 $ spatial
universes to sufficiently suppress the fluctuations around each of the mean
values such as to be able to distinguish the two phases. On such large
lattices, one indeed observes the characteristic double-peak structure in a
histogram of the action per lattice link at $c=0.2359$,
cf.~the right-hand panel of fig.~\ref{fig2359}.
On smaller lattices, the fluctuations in each individual phase swamp the
difference in the mean action per lattice link between the phases and
the double-peak structure cannot be resolved. It should be noted
that the weakness of the transition obtained near the physical point is a
nontrivial prediction of the model; in fact, at other, nonphysical points
in the plane of coupling constants, one can observe very strongly first order
phase transition behaviour, cf.~fig.~\ref{small_histogram}.
This figure was obtained at $c=0, \epsilon = 0.993$ on very small,
$4^3 \times 2$ lattices. Even then, the very large difference in the action
density between the two phases is clearly visible. Note that this case (with
no curvature action, $c=0$) corresponds to (the dual formulation of) standard
$Z(3)$ lattice gauge theory.

The weakness of the transition observed near the physical point in
figs.~\ref{fig2359} and \ref{fig2357} matches the result in full $SU(3)$
Yang-Mills theory, where the first order character of the deconfining
phase transition also turns out to be rather weak \cite{weak}.
This constitutes a successful nontrivial test of the correspondence
between $SU(3)$ Yang-Mills theory and the present random vortex
world-surface model.

\begin{figure}[t]
\centerline{
\begin{minipage}{7.5cm}
\includegraphics[width = 7.5cm]{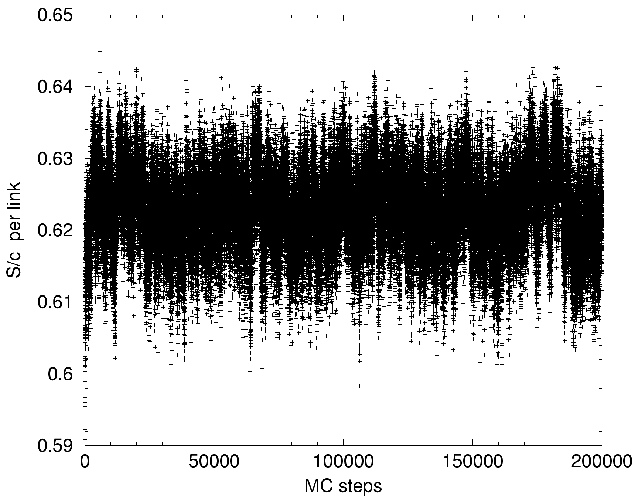} \end{minipage}\hfill 
\begin{minipage}{7.5cm}
\includegraphics[width = 7.5cm]{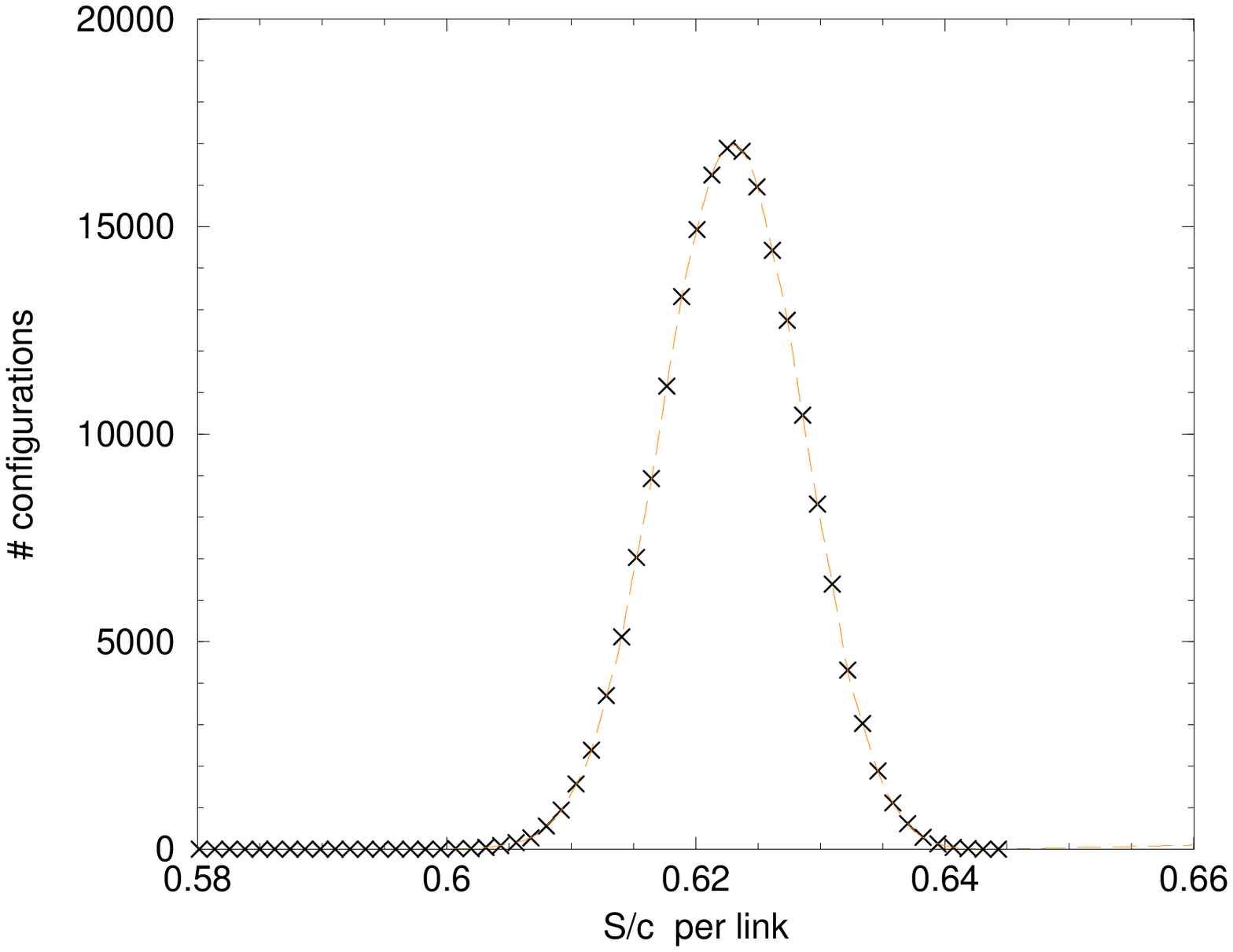}\end{minipage}
}
\caption{Left-hand panel: Action per link as a function of Monte Carlo time in
the region of the deconfining phase transition for the $SU(2)$ vortex model
studied in \cite{R1}. Measurements were taken on a $30^3 \times 2$ lattice
with $\epsilon = 0$ and $c = 0.2990$. This data should be compared to the
$SU(3)$ case depicted in fig.~\ref{fig2359}. At least to the accuracy which was
sufficient to detect a double peak structure in the $SU(3)$ action
distribution, the $SU(2)$ model exhibits the characteristics of a second
order transition, as shown by the single peak structure of the $SU(2)$
action distribution displayed in the right-hand panel.}
\label{fig2990}
\end{figure}

To complete the discussion of the order of the deconfining phase
transition in the vortex model, the authors also revisited the $SU(2)$
case, in which the transition is expected to be second order. This
topic was not studied in detail in \cite{R1}. As above, $30^3 \times 2$
lattices were used and the value of $c$ at which the phase transition
occurs was localized in the interval $0.298 < c < 0.300$ by looking for
the maximal slope in the curvature action per lattice link. Taking the
$SU(3)$ case as an indication, where the first order character of the phase
transition was discernible over a range of $c$ of width larger than
$0.0002$, cf.~figs.~\ref{fig2359} and \ref{fig2357}, the authors scanned the
region $c=0.296$ to $c=0.302$ in steps of $0.0002$ and recorded Monte Carlo
histories analogous to figs.~\ref{fig2359} and \ref{fig2357}. None of them
displayed a significant indication of a first order phase transition
(cf.~fig.~\ref{fig2990} as an example; the other Monte Carlo
histories were of the same qualitative character). Thus, at least to the
accuracy which proved sufficient to ascertain the order of the transition
in the $SU(3)$ case, the $SU(2)$ model appears to behave in a manner
which is consistent with $SU(2)$ lattice Yang-Mills theory and which is
expected from Ginzburg-Landau analysis \cite{svetitsky}, i.e., it appears
to display a second-order deconfinement phase transition. Of course, the
data do not exclude a first order transition of considerably weaker character
than in the $SU(3)$ case. The authors did carry out a more detailed finite-size
analysis using lattices of varying spatial extension.

\section{The structure of vortex clusters}
\label{cluster}

The confinement properties of the model are intimately tied to the
percolation properties of the vortices. In the $SU(2)$ case studied in
ref.~\cite{R1}, a \emph{percolation} transition at finite
temperatures, inducing deconfinement, became apparent. This analysis can be
analogously extended to the present $SU(3)$ version of
the model. Since the findings are very similar to the aforementioned $SU(2)$
case, this section will be kept rather brief. For a detailed exposition of 
the connection between vortex percolation and confinement, the reader is
referred to \cite{R1}.\footnote{It should be noted that the center projection
vortex ensemble extracted from full $SU(2)$ lattice Yang-Mills theory exhibits
the same percolation mechanism for the deconfinement phase transition
\cite{R14} as the present random vortex world-surface model.}

In order to exhibit the percolation properties of vortices, it is useful to
consider three-dimensional slices of space-time, taken {\em between} vortex
lattice (hyper-)planes, keeping either one space coordinate or the time
coordinate fixed. Since this situation will appear frequently in the
following, such three-dimensional slices will be referred to as \emph{space}
and \emph{time slices}, respectively. 

\begin{figure}[t]
\centerline{
\begin{minipage}{15cm}
\begin{minipage}{4.8cm}
\includegraphics[width = 4.8cm]{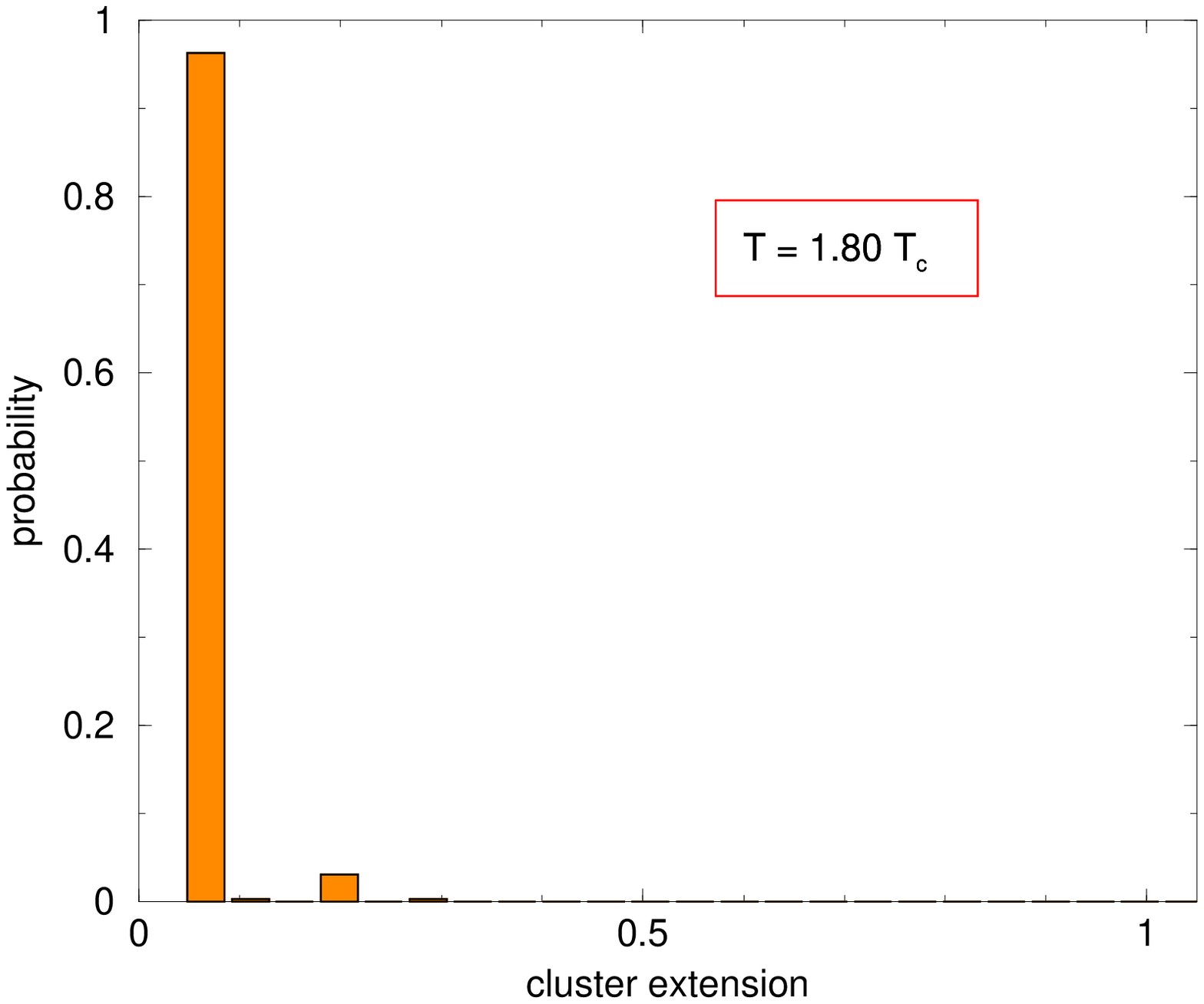} 
\end{minipage}
\hfill 
\begin{minipage}{4.8cm}
\includegraphics[width = 4.8cm]{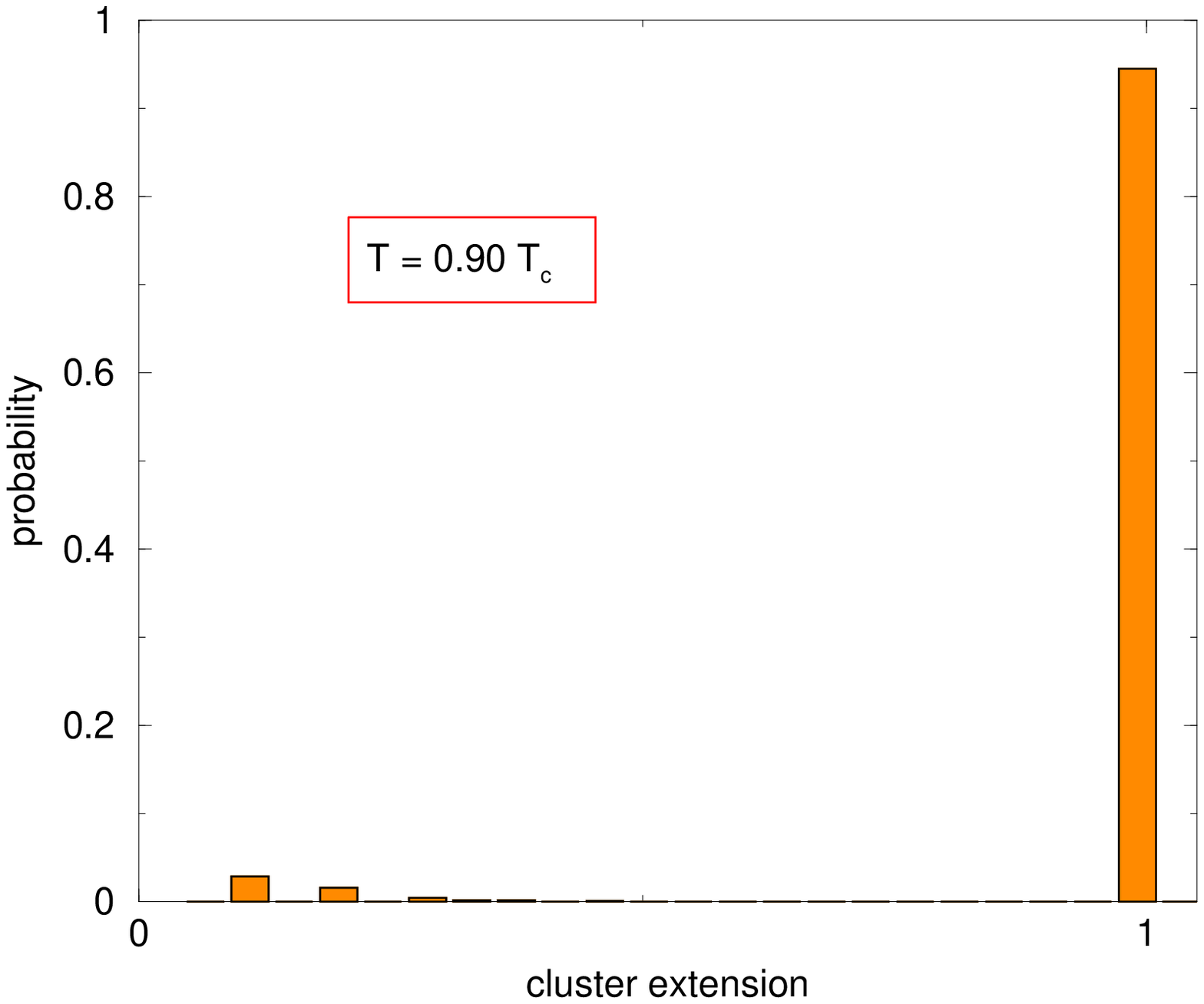}
\end{minipage}
\hfill 
\begin{minipage}{4.8cm}
\includegraphics[width = 4.8cm]{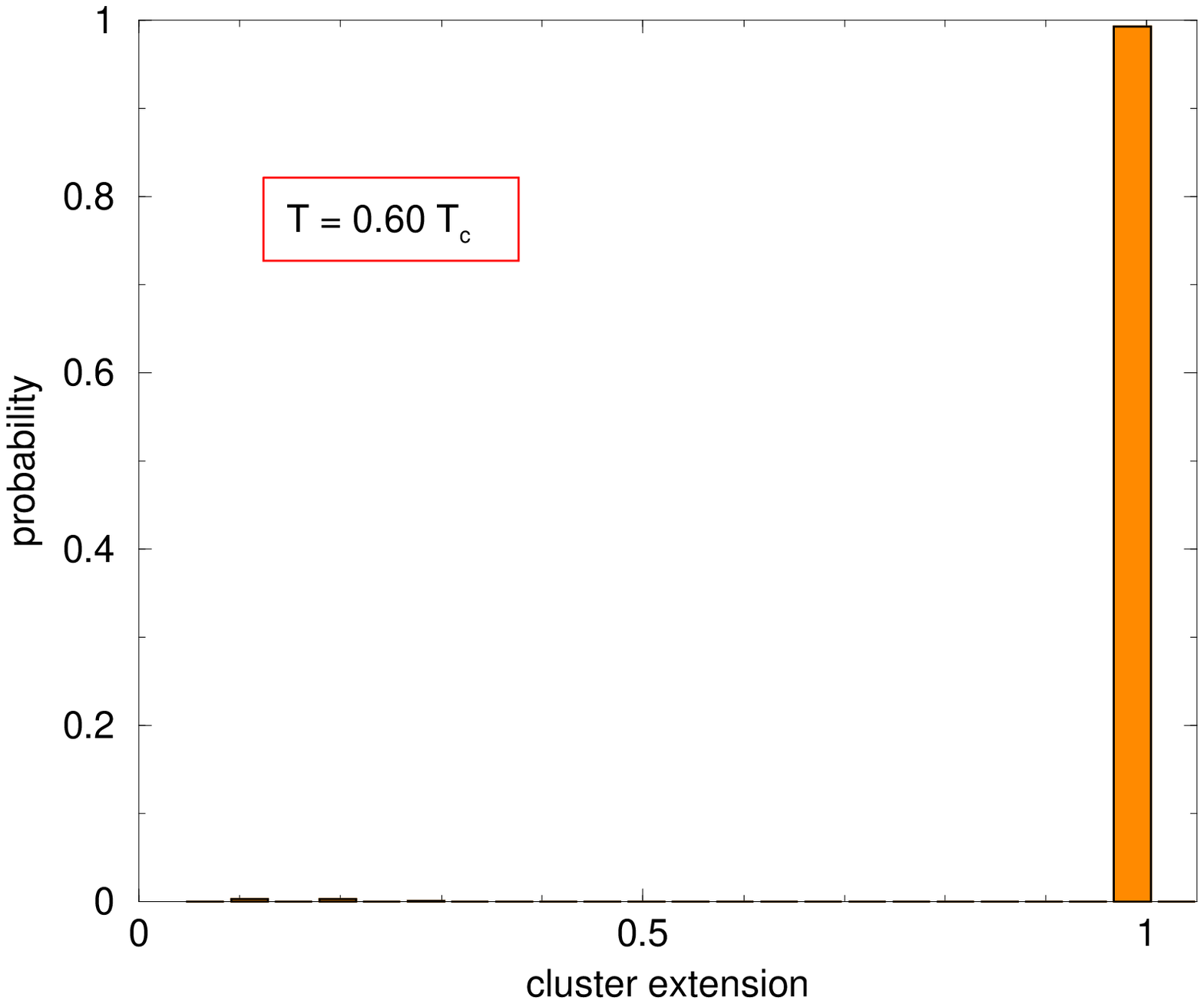}
\end{minipage}
\end{minipage} 
}
\centerline{
\begin{minipage}{15cm}
\begin{minipage}{4.8cm}
\includegraphics[width = 4.8cm]{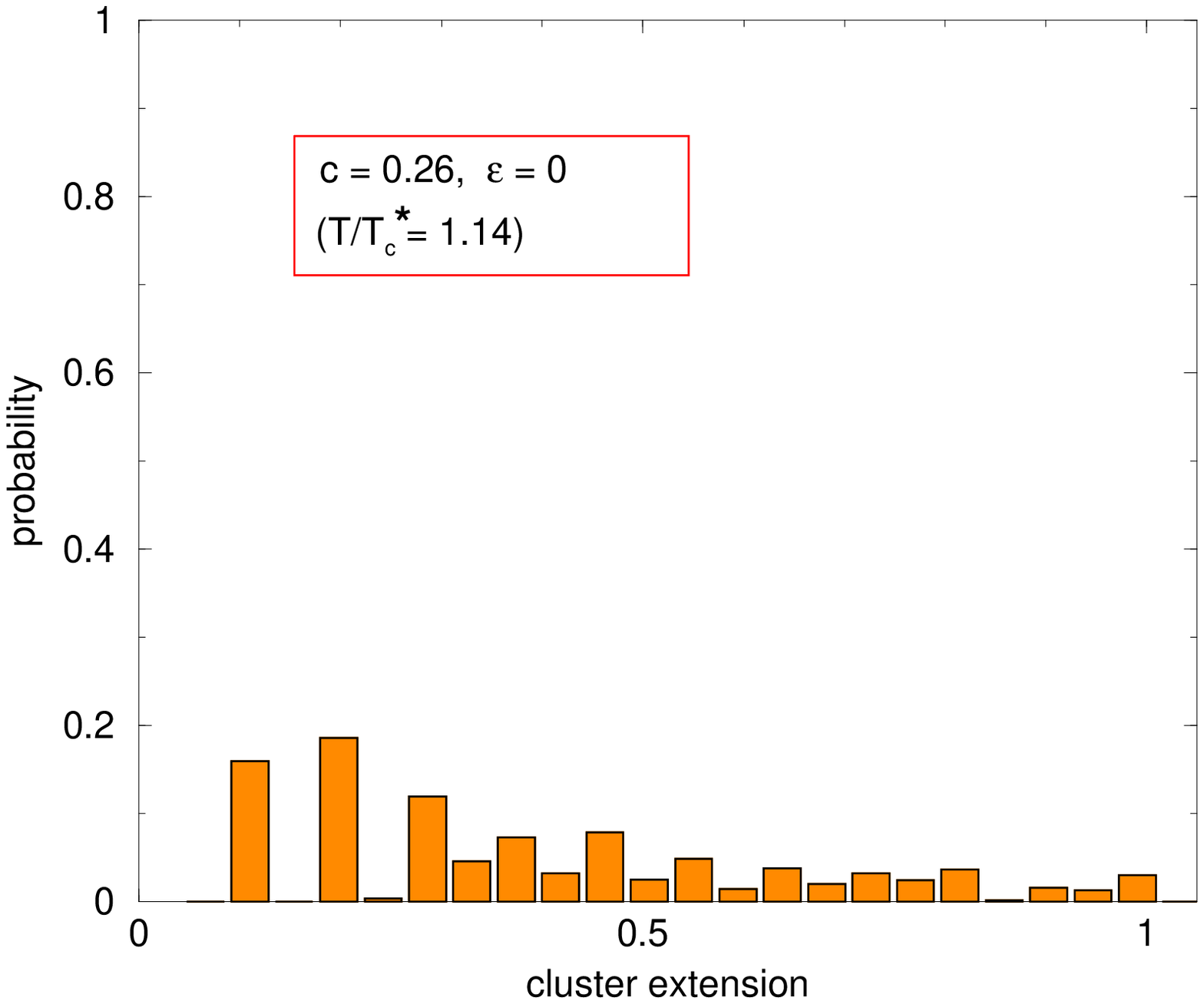} 
\end{minipage}
\hfill
\begin{minipage}{4.8cm}
\includegraphics[width = 4.8cm]{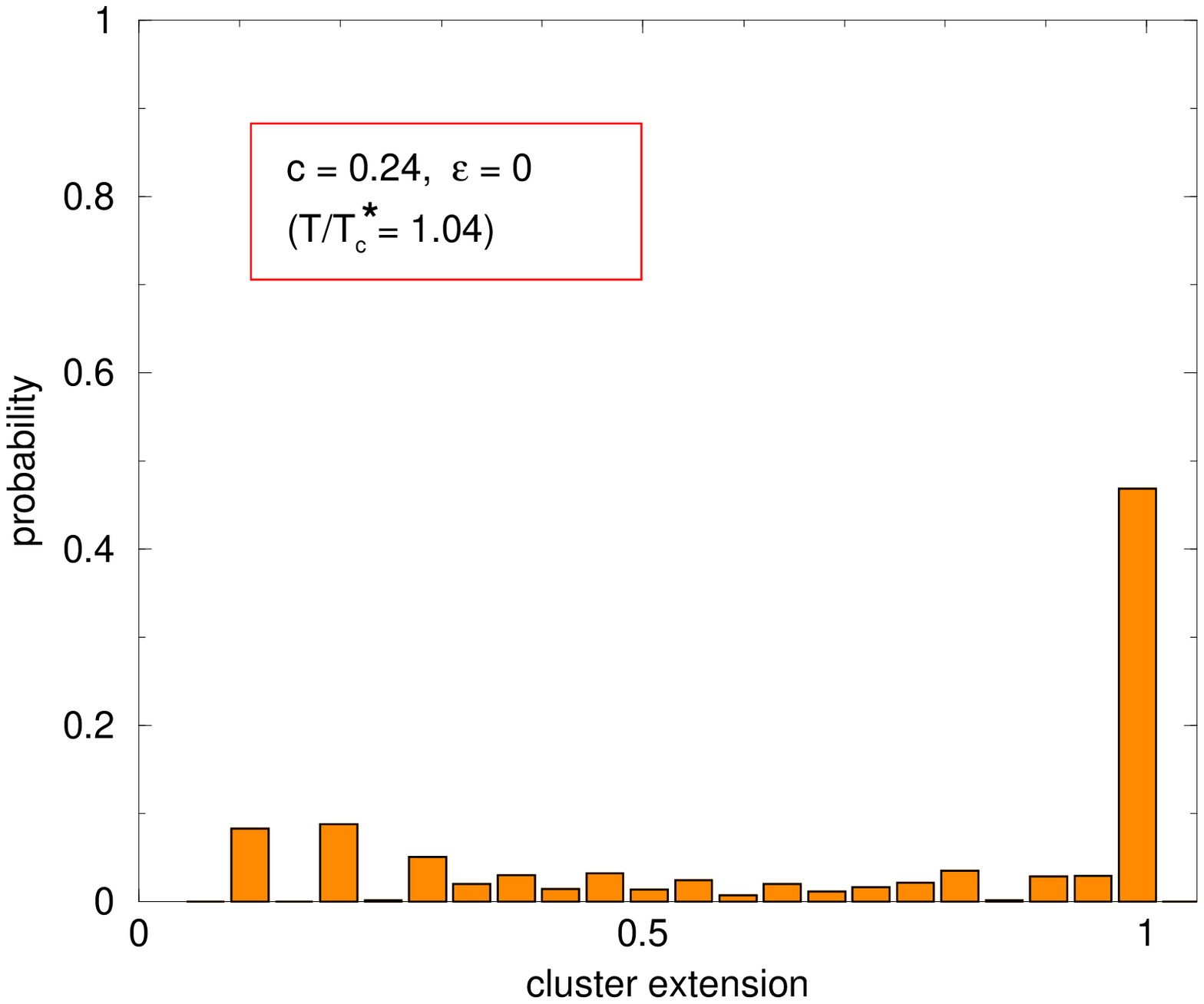}
\end{minipage}
\hfill
\begin{minipage}{4.8cm}
\includegraphics[width = 4.8cm]{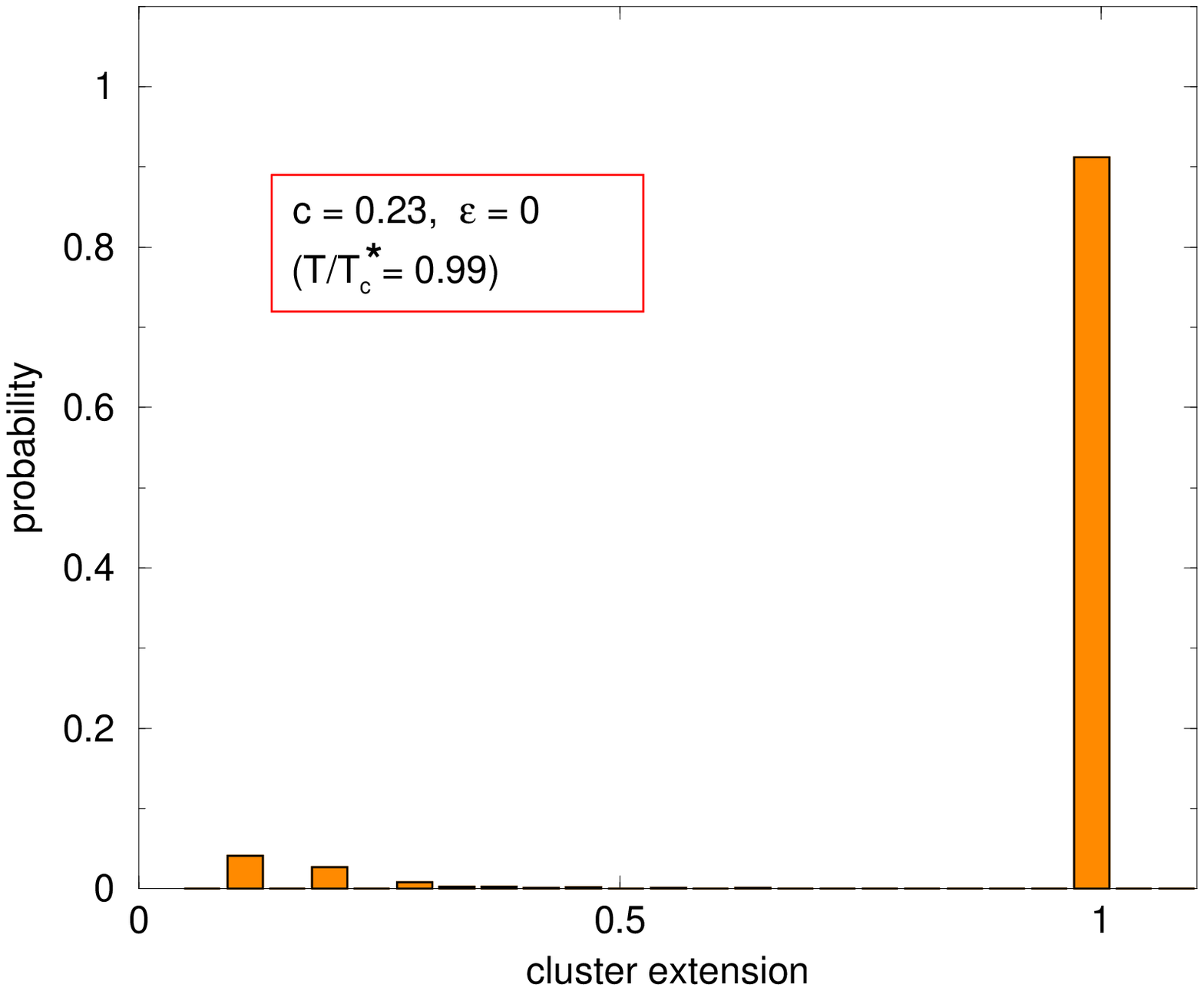}
\end{minipage}
\end{minipage}
} 
\caption{Probability distributions obtained by binning vortex links belonging
to space slices of the random vortex surfaces according to the extension
of the vortex cluster they belong to. Cluster extensions are measured in
units of the maximal possible extension, see main text. The top line shows 
measurements taken using $16^3 \times N_0 $ lattices at the physical point
(\ref{physical}), both in the 
confined and the deconfined phase. Since these graphs do not give
a detailed picture of the behaviour close to the phase transition itself, 
the lower line presents cluster distributions taken on a $16^3 \times 2$ 
lattice with $\epsilon=0$ and $c = 0.26,\,0.24,\,0.23$ (from left to 
right). While these points correspond formally to 
$T / T_c^\ast = 1.14,\,1.04\,,0.99$, they do not lie on the physical 
trajectory. As a consequence, the transition temperature $T_c^\ast$ 
at these points (as obtained by the interpolation procedure described in
section \ref{coupling}) is \emph{not} identical to the physical $T_c$, but
rather a formal quantity which is undetermined in absolute units. As already
explained for the Monte Carlo histories in the previous section, the
parameters are still sufficiently close to the physical trajectory to provide
an accurate indication of physics there.} 
\label{clust_ext}
\end{figure}

In any such three-dimensional slice, vortices form closed loops made up of
links which, in general, will self-intersect to form complicated
vortex \emph{clusters}. A cluster is defined as a (maximal) set of 
connected vortex links and its extension as the maximal Euclidean 
distance between any two links in that cluster. Fig.~\ref{clust_ext} 
shows the probability distribution for vortex links to belong to a cluster 
of a given extension. These measurements were taken in space slices
and exhibit a clear signal of a percolation transition: 
Below $T_c$, most vortex links are in clusters of nearly maximal 
extension,\footnote{Due to the periodic boundary conditions, the maximal 
possible cluster extension on a $N_0 \times N_s^3$ lattice is 
$\sqrt{N_0^2 + 2 N_s^2}\,a/2$ for space slices and $\sqrt{3} N_s a/2$ for
time slices.} while the deconfined region above $T_c$ is dominated by
many small clusters which do not percolate. Note that the persistence
of clusters of maximal extension slightly above the deconfinement
temperature (cf.~center panel in the lower line of fig.~\ref{clust_ext})
is natural. At temperatures so close to the phase transition, one must
expect a significant density of clusters which are as large as the lattice
universe used, and which would only be revealed as non-percolating in
significantly larger lattice universes. To verify this, a detailed
finite-size analysis using lattices of different extensions is necessary.
In complete correspondence, of course, the deconfinement temperature itself
is only defined up to finite-size effects.
  
\begin{figure}[t]
\centerline{
\begin{minipage}{4.8cm}
\includegraphics[width = 4.8cm]{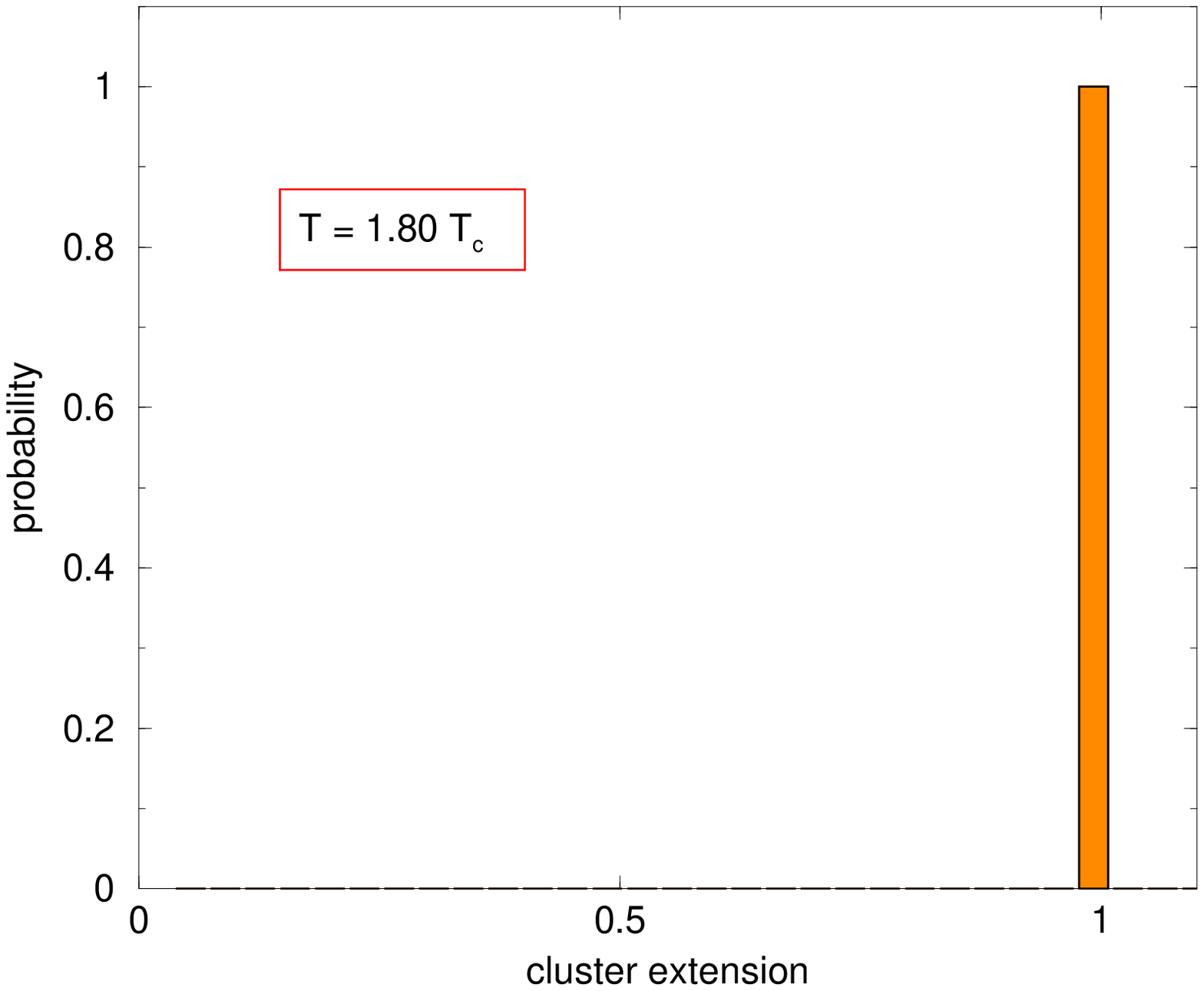} 
\end{minipage}
\hfill
\begin{minipage}{4.8cm}
\includegraphics[width = 4.8cm]{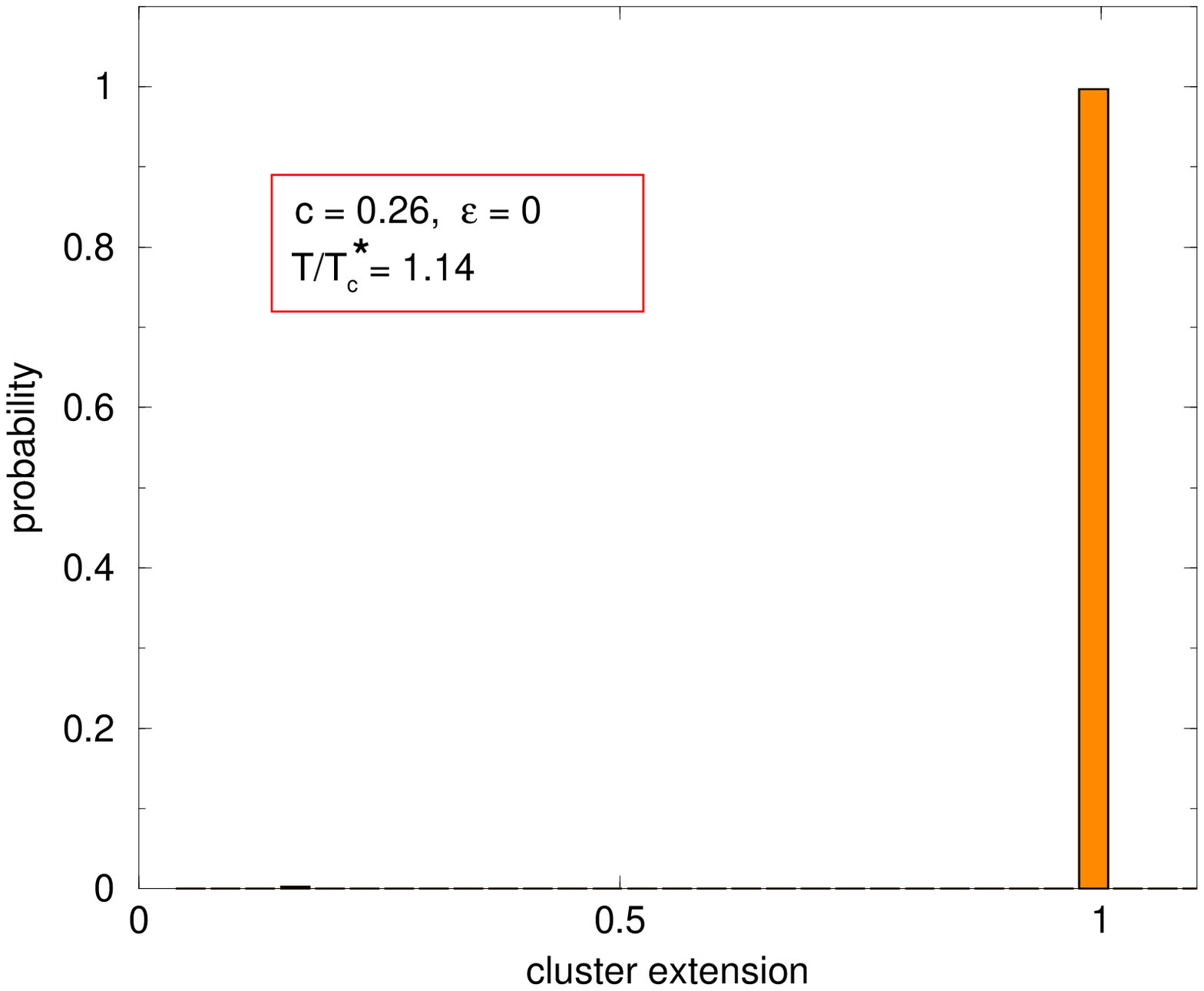} 
\end{minipage}
\hfill
\begin{minipage}{4.8cm}
\includegraphics[width = 4.8cm]{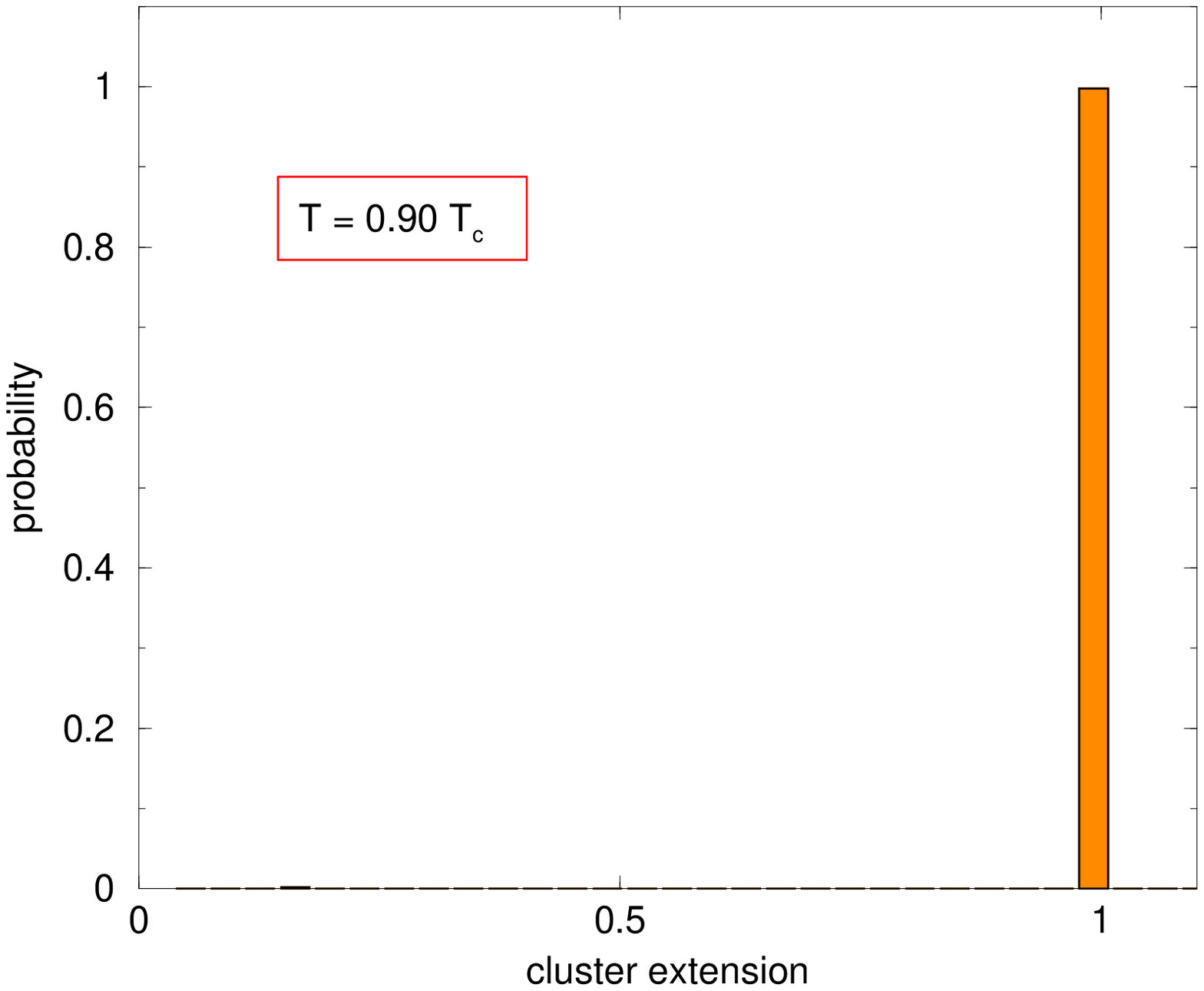} 
\end{minipage}
}
\caption{Probability distributions obtained by binning vortex links belonging
to time slices of the random vortex surfaces according to the extension
of the vortex cluster they belong to. Cluster extensions are measured in units
of the maximal possible extension, see main text. While the left and
right-hand histograms were obtained using $16^3 \times N_0 $ lattices at
the physical point $(c=0.21,\epsilon =0)$, the graph in the middle results when
using a $16^3 \times 2$ lattice at the point $(c=0.26,\epsilon =0)$, which
does not lie on the physical trajectory. In analogy to the discussion
in the caption of fig.~\ref{clust_ext}, the label $T / T_c^\ast = 1.14$
indicates that the measurement is
performed slightly above the phase transition temperature, but the "local"
$T_c^\ast$ at an unphysical point on the plane of coupling constants is a 
formal quantity and not identical to the physical $T_c=277$ MeV in the 
left and right-hand panels.}
\label{clust_ext3}
\end{figure}

The above picture changes drastically if one considers time slices
instead of space slices. Here, vortex lines percolate in \emph{both} phases.
This is seen in the probability distribution of 
vortex cluster extensions in time slices. To compare with the 
previous results in fig.~\ref{clust_ext}, one now has to normalise
to the maximal possible extension in time slices, i.e.~$\sqrt{3} N_s a/2$.
As can be seen from fig.~\ref{clust_ext3}, the distributions show no 
sign of a phase transition and remain strongly peaked at the maximal
extension for all temperatures. This means that virtually all vortex links
in time slices belong to clusters of maximal spatial extension; the clusters
percolate for all temperatures. 

A more detailed picture of the deconfined phase emerges from the study of
the number of links contained in the clusters. For the
set of parameters (\ref{physical}), only $N_0 = 1$ realises the 
deconfined phase. In space slices of this lattice, $96.5$\% of all
vortex links are contained in clusters with only one link (those
links \emph{are} the entire cluster and wind directly around the Euclidean 
time direction). At $N_0 = 2$, which is below $T_c$, the fraction of 
vortex links belonging to winding clusters consisting of two links is only
about $2.9$\%, while most vortex flux is contained in clusters of roughly 
$500$ links. For time slices, on the other hand, there is no significant
difference between $N_0 = 1$ and $N_0 = 2$: The majority of vortex links
in both cases is contained in clusters made up of more than $4000$
links. Thus, the small vortex clusters which dominate the deconfined 
phase in space slices are mainly winding vortex configurations which run
directly along the compactified time direction.
 
The situation is different at (unphysical) points
in coupling constant space where deconfinement is observed even at zero 
temperature. In this case, there is of course no finite temperature phase
transition and analogous measurements show that no percolating vortex
clusters exist in either time or space slices, at all temperatures. All
these findings are virtually identical to the corresponding results
obtained for the $SU(2)$ case in \cite{R1}. To summarise, this confirms that  
also in the $SU(3)$ case, confinement is generated by
percolating vortex clusters while the deconfined phase is characterised 
(in space slices) by small disconnected clusters which predominantly 
wind around the (short) Euclidean time direction of the lattice universe.
For a more detailed physical discussion of the connection between vortex
percolation and confinement, the reader is referred to \cite{R1}.


\section{Vortex branching}
\label{split}
The model action, eqs.~(\ref{area}) and (\ref{curvature}), presented in
section \ref{sec:2} is essentially equivalent to the $SU(2)$ case \cite{R1}.
It does not contain terms that enhance or penalise vortex 
branchings \emph{explicitly}. Nonetheless, the finite temperature studies in 
section \ref{phase} indicate that the larger phase space of $SU(3)$ vortices,
i.e. the possibility of vortex branchings,  can make a physical difference to 
the $SU(2)$ case: It changes the order of the deconfinement phase transition. 
It is therefore interesting to study the distribution of vortex 
branchings in the Monte Carlo ensembles in more detail.

To do so, it is useful to consider three-dimensional slices of space-time as
in the previous section. Vortices branch along lines (links) in four dimensions,
so taking a three-dimensional slice yields a three-dimensional distribution of
\emph{branching points}. Just as any given link in four dimensions is attached
to six elementary squares (which may or may not be part of a vortex), each site
in the three-dimensional slice can be attached to up to six vortex links. As
explained in section \ref{sec:2}, the model description does not keep track of
vortex orientation and cannot distinguish between, say, a $z_2$-vortex
branching into two $z_1$-vortices, or three $z_1$-vortices annihilating along
a common link in four dimensions. In fact, corresponding to the symmetry of
the underlying action with respect to the two possible types of vortex flux,
the only measurable physical quantity is the \emph{number} $\nu$ of vortex
surfaces meeting at each link,
%
%

\begin{itemize}
\item $\nu = 0$ indicates that the link is not part of any vortex
\item $\nu = 1$ is forbidden, since vortices are closed and there are
              no end links (Bianchi's identity) 
\item $\nu = 2$ indicates that the link is part of a vortex surface which
                does not branch or self-intersect at that link
\item $\nu = 4,6$ represent vortex self-intersections already present
                  in the $SU(2)$ case  
\item $\nu = 3,5$ indicate true $SU(3)$ vortex branchings which have no 
                   counterpart in $SU(2)$
\end{itemize}

\begin{figure}[t]
\centerline{
\begin{minipage}{5cm}
\includegraphics[width = 5cm]{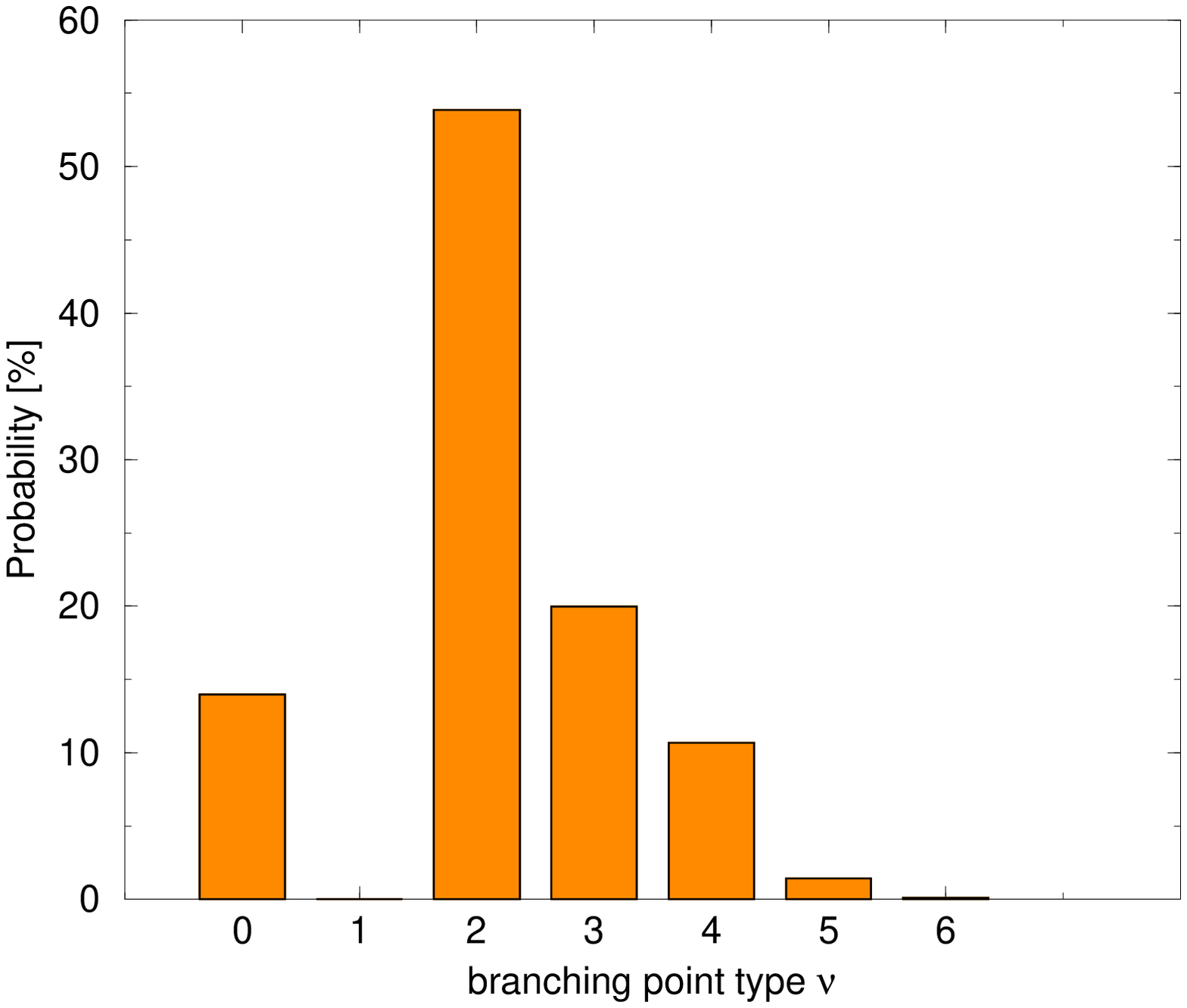}\end{minipage}\hfill 
\begin{minipage}{5cm}
\includegraphics[width = 5cm]{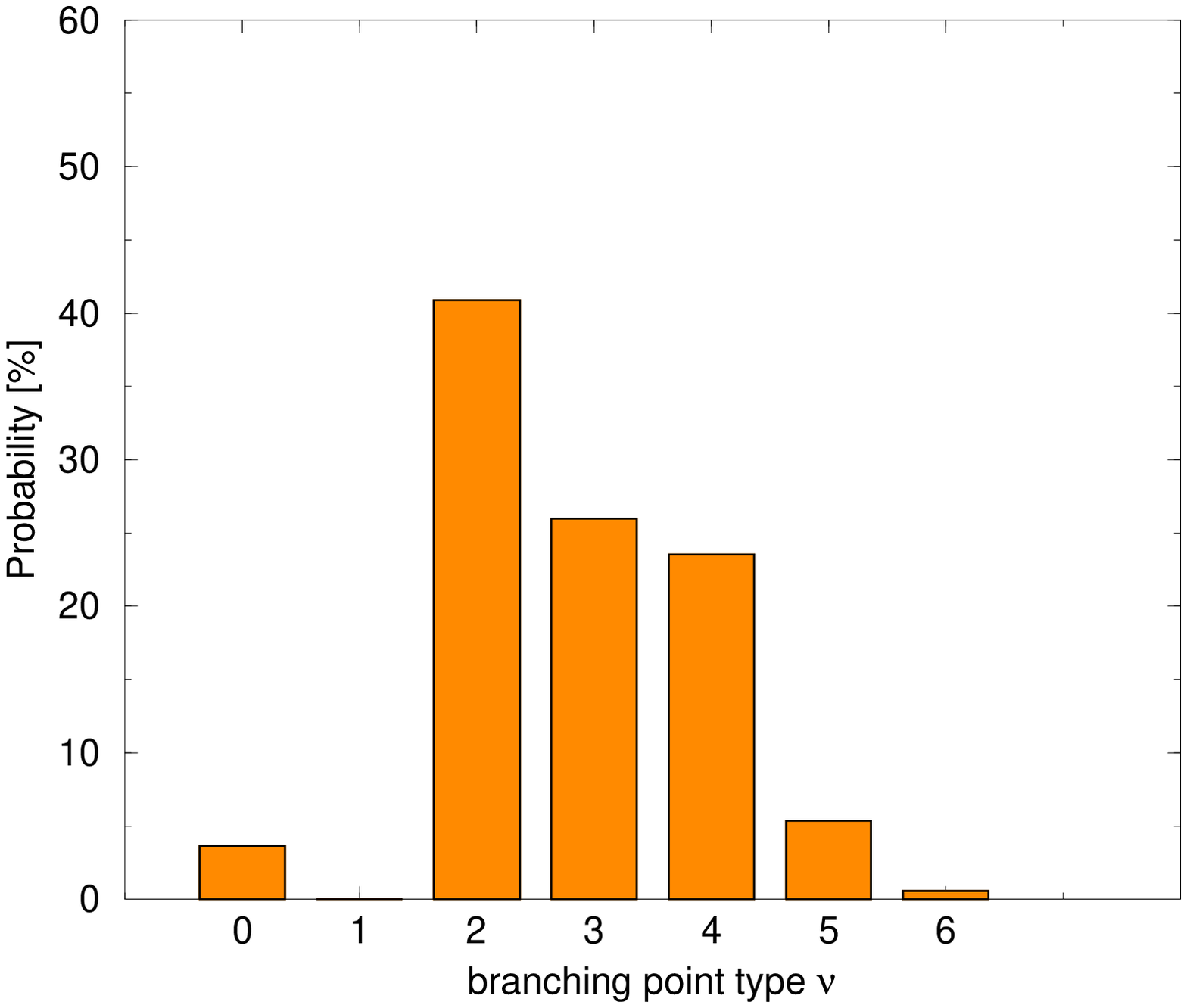}\end{minipage}\hfill
\begin{minipage}{5cm}
\includegraphics[width = 5cm]{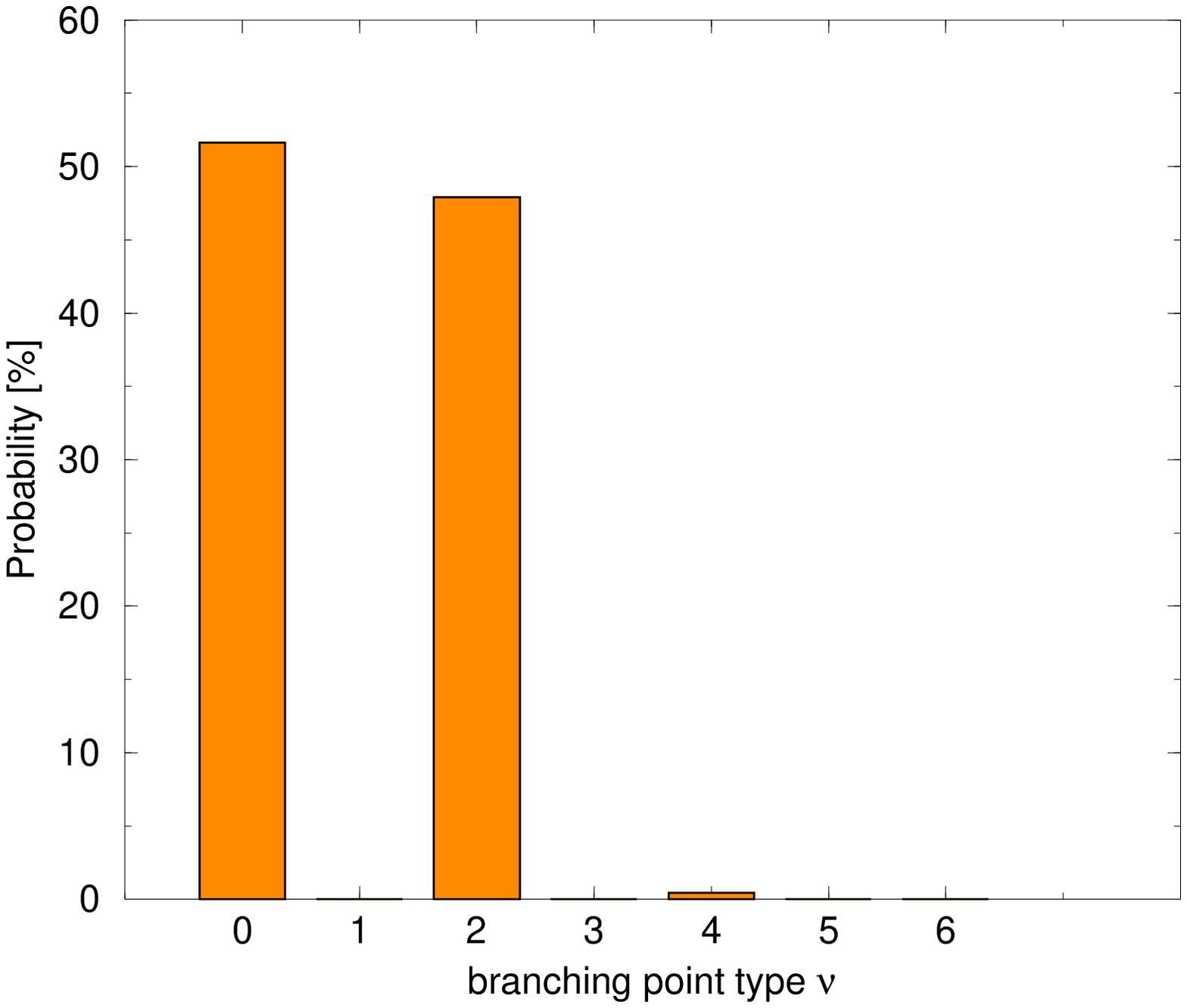}\end{minipage}
}
\caption{Distribution of branching points in three-dimensional slices of
space-time, all measured at the physical point $\epsilon = 0$ and $c= 0.21$.
The left-hand panel shows the results obtained using a symmetric $16^4$
lattice (i.e.~at zero temperature) while the center and right-hand panels
give the distributions obtained using a $16^3 \times 1$ lattice (i.e.~in
the deconfined phase). The graph in the middle refers to a \emph{time slice}
while the right-hand panel shows the corresponding result for a
\emph{space slice}.}
\label{split1}
\end{figure}

An example of a branching with $\nu =3$ is displayed in fig.~\ref{branch_fig}.
Figure \ref{split1} shows the result of measurements of the $\nu$ distribution
in three-dimensional slices of space-time at the physical 
point, eq.~(\ref{physical}). In all cases, there are indeed no vortex 
end points, $\nu = 1$, as vortices are constrained to be closed. For the 
symmetric $16^4$ lattice used to obtain the left-hand panel, the majority of
points is associated with $\nu = 2$, i.e.~they represent sites where the
vortex does not branch. However, vortex branchings $\nu=3,5$ as well as
self-intersections $\nu=4,6$ occur with a significant probability, indicating
that the structure of vortices is quite fibrated in this case. Note that a 
sufficiently large symmetric lattice is approximately at $T=0$, which 
is in the confining phase for the parameters eq.~(\ref{physical}).

The two other panels of fig.~\ref{split1} show the branching point
distribution for the same parameters, eq.~(\ref{physical}), but now at high 
temperatures (on an asymmetric $16^3 \times 1$ lattice) where the 
model is in the deconfined phase. For \emph{time slices} 
(center panel), there is no qualitative change 
as the probability of branchings even increases slightly. This indicates 
that vortex fibration is significant along temporal links and the structure 
of vortex clusters in time slices does not change qualitatively across 
the phase transition. This result is in line with the earlier finding
that vortex percolation persists in time slices in the deconfined phase.

In \emph{space slices} (cf.~right-hand panel of fig.~\ref{split1}), the
vortex structure changes strongly as one increases the temperature into
the deconfined regime. Branching points and self-intersections are strongly
suppressed, even though vortices still abound; almost half of the lattice
sites in a three-dimensional space slice are still located on vortices.
This is consistent with the fact that the vortices are to a large extent
aligned along the short Euclidean time direction;
branchings, which would imply vortices splitting off into spatial
directions, are correspondingly rare.
In a space slice, the complicated, fibrated vortex cluster of the 
confining phase has turned into small, mostly non-fibrated vortices which 
are not interconnected and cease to percolate. It should be noted that 
the fraction of lattice sites in the three-dimensional space slice with 
$\nu \neq 0$ drops only moderately across the phase transition -- from 
$85 \%
$ in the confining phase to $50 \% 
$ in the deconfined phase. 
Thus, it is really the change in vortex structure engendered by the
percolation transition, not the overall vortex density, which drives
the deconfining phase transition. 

\begin{figure}[t]
\centerline{
\begin{minipage}{5cm}
\includegraphics[width = 5cm]{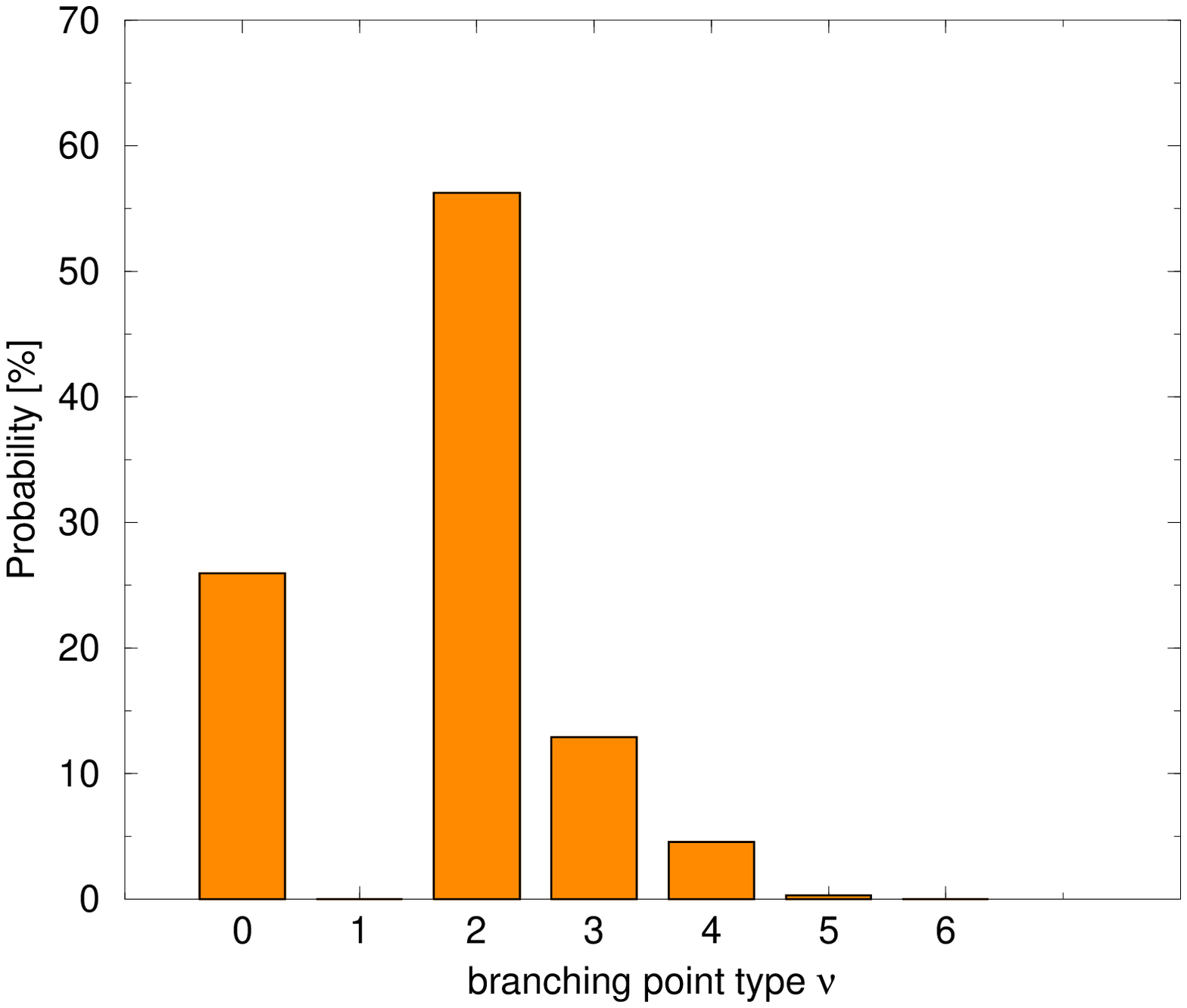}\end{minipage}\hfill 
\begin{minipage}{5cm}
\includegraphics[width = 5cm]{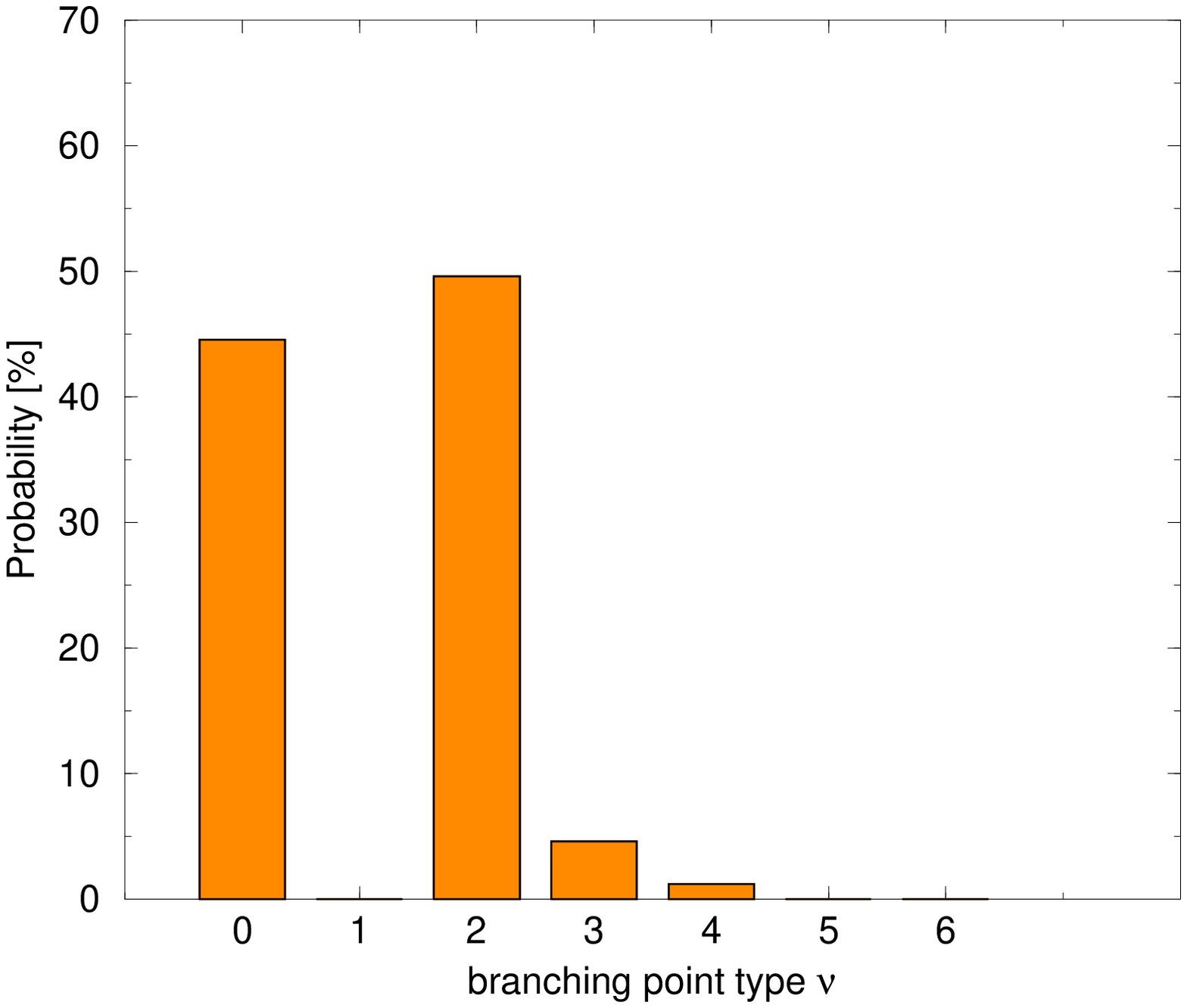}\end{minipage}\hfill
\begin{minipage}{5cm}
\includegraphics[width = 5cm]{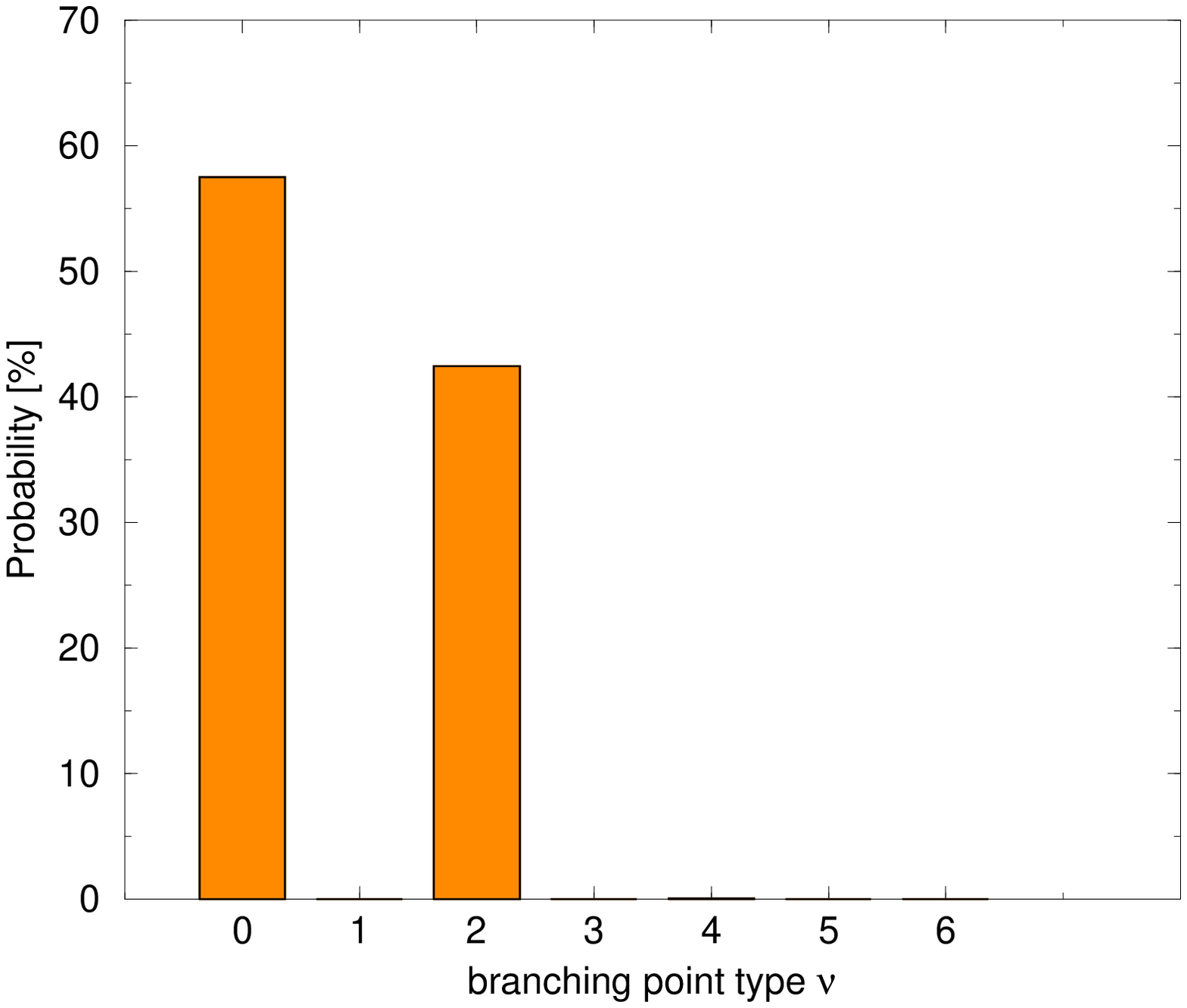}\end{minipage}
}
\caption{Distribution of branching points in three-dimensional space slices
obtained at the (unphysical) point $\epsilon = 0$ and $c= 0.30$. The 
measurements were taken on an asymmetric $16^3 \times N_0$ lattice, with 
$N_0 = 3,2,1$ (from left to right), i.e.~the temperature increases
from left to right. The distribution at $N_0 = 3$ (left) is virtually 
identical to the zero-temperature (confining) case $N_0 = 16$, while
$N_0 = 2$ (center) and $N_0 = 1$ (right) are in the deconfined phase.}
\label{split3}
\end{figure}

For the physical point $\epsilon = 0$, $c=0.21$, it is necessary to let the 
Euclidean time extension of the lattice universe shrink to $N_0 = 1$ in order
to observe the deconfined phase. This case is, however, somewhat special,
since then, vortices extending in the time direction necessarily close
via the periodic boundary conditions, i.e.~they automatically wind around 
the Euclidean time dimension. To corroborate that the suppression of 
branching points in the deconfined phase is \emph{not} an artefact of
$N_0 = 1$, the authors have furthermore investigated the (unphysical) point
$\epsilon = 0$, $c=0.30$, where $N_0 = 2$ is also well within that
phase. Fig.~\ref{split3} shows data measured on $16^3 \times N_0$ 
lattices, for various values of $N_0$, i.e.~various temperatures. The 
change in vortex structure discussed above, i.e., the suppression of
branchings, now also becomes apparent in the $N_0 = 2$ case, although
it should be noted that the branching density evidently does not behave
as a strict order parameter; while strongly suppressed, the probability
of branchings displayed in the center panel of fig.~\ref{split3} is not
entirely negligible.

\begin{figure}[t]
\centerline{
\begin{minipage}{7cm}
\includegraphics[width = 7cm]{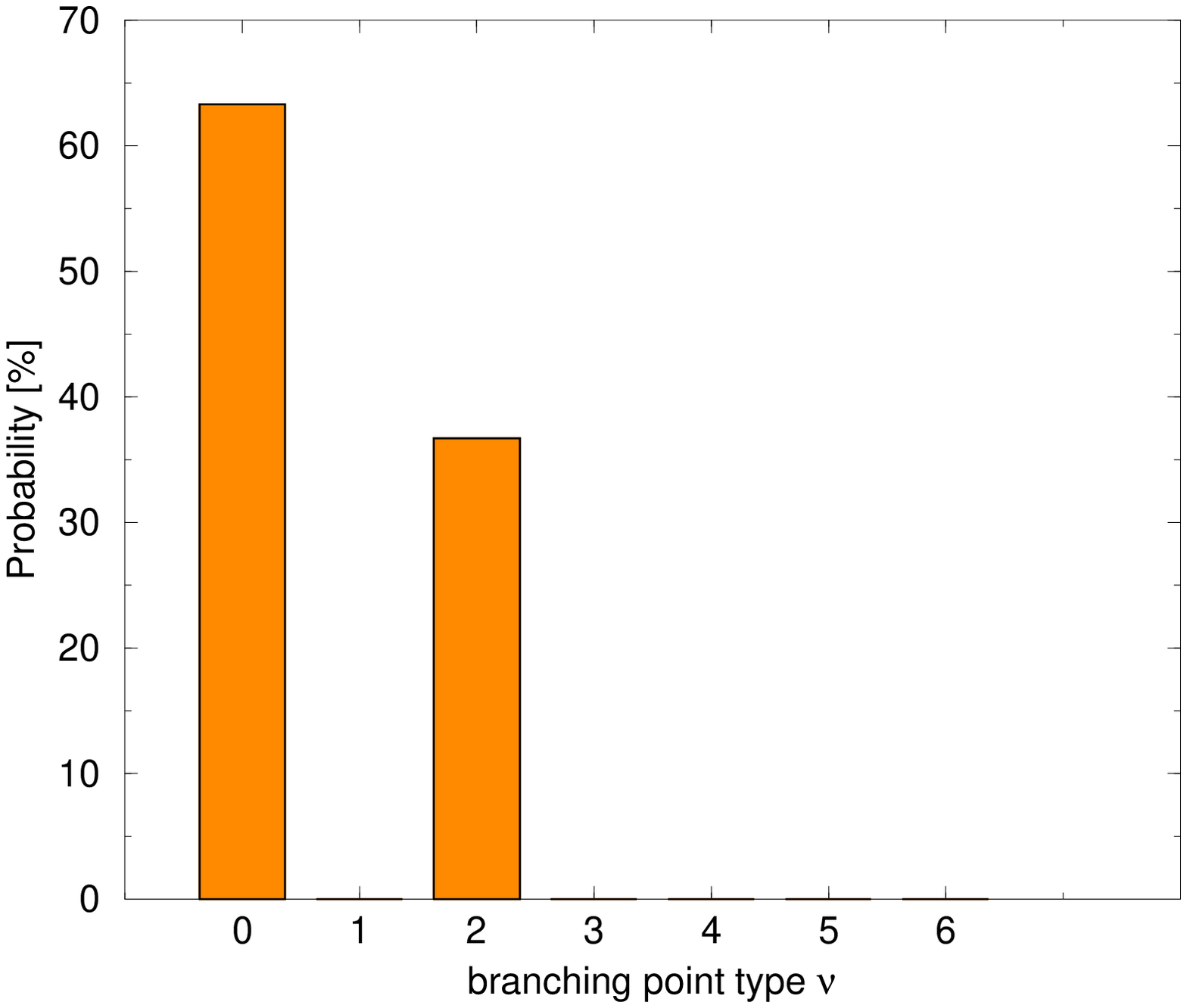}\end{minipage}\hfill 
\begin{minipage}{7cm}
\includegraphics[width = 7cm]{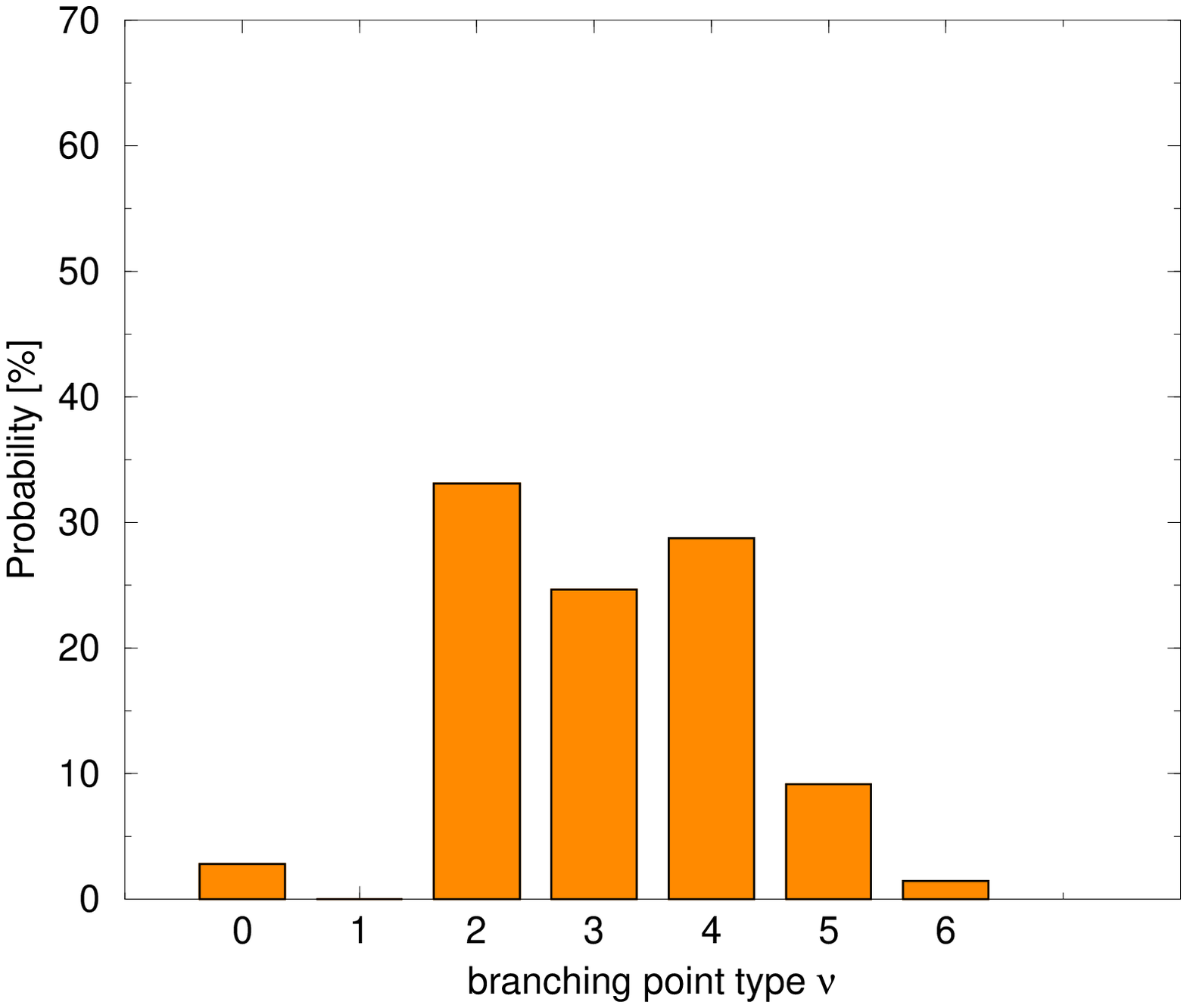}\end{minipage}
}
\caption{Distribution of branching points in three-dimensional slices of
space-time measured using a $16^4$ lattice at the points $\epsilon = 0.25$,
$c = 0.3$ (deconfined region, left-hand panel) and $\epsilon = -0.2$,
$c = 0.1$ (confining region, right-hand panel).}
\label{split2}
\end{figure}

Similar results are also seen as one traverses the crossover  
in fig.~\ref{couplingspace} by varying the coupling constants $\epsilon$
and $c$. Figure \ref{split2} presents the distribution of branching points
in three-dimensional slices of space-time measured using a symmetric $16^4$
lattice, both at the point $\epsilon = 0.25$, $c = 0.3$ in the deconfined
region (left-hand panel) and at $\epsilon = -0.2$, $c = 0.1$ in the confining
region (right-hand panel). While branching points are abundant in the confining
region (indicating the existence of heavily fibrated vortex clusters in 
this case), they are virtually absent in the deconfined region. 
Since there are no short directions on a $16^4$ lattice for vortices to 
wind around, the absence of branching points for the deconfining parameters 
in this case is due to space-time being filled with small vortex structures
which are disconnected and do not percolate in \emph{any} direction. 

\section{Conclusions and outlook}
\label{conclusions}
In the present work, the random vortex world-surface model for the infrared
sector of Yang-Mills theory, previously developed for the gauge group $SU(2)$
\cite{R1}, was extended to the gauge group $SU(3)$. As in the $SU(2)$ case,
it proved possible to quantitatively reproduce the confinement properties
of the corresponding lattice Yang-Mills theory. Both the low-temperature
confining phase as well as the high-temperature deconfined phase are
encompassed by the model, and its coupling constants can be chosen such
that the ratio $T_c /\sqrt{\sigma_{0} } \approx 0.63$ of full Yang-Mills
theory is recovered.

Having fixed the parameters of the model, an accurate prediction of the
spatial string tension in the deconfined phase is obtained, as in the
$SU(2)$ case. Furthermore, the deconfinement phase transition is predicted
to be {\em weakly } first order, in agreement with $SU(3)$ lattice gauge
theory.

The confinement properties of the model are intimately tied to the percolation
properties of the vortices. As in the $SU(2)$ case, confinement is generated
when vortices in space slices of the lattice universe (i.e., keeping one
spatial coordinate fixed) percolate; by contrast, the deconfined phase is
characterized by small, mutually disconnected vortices which are predominantly
aligned with the short Euclidean time direction of the space slice and are
closed by virtue of the periodic boundary conditions. Nevertheless, there is
one important difference in geometry between the $SU(2)$ and $SU(3)$
cases: In the latter case, there are two distinct types of quantized vortex
flux and, as a consequence, vortices can branch. This does not happen for
$SU(2)$ vortices. In particular, when traversing the finite-temperature phase
transition into the deconfined phase, vortex branchings observed in space
slices of the lattice universe become strongly suppressed, consistent with the
aforementioned
alignment of the vortices with the Euclidean time direction in that phase.
Note that a significant presence of branchings is only possible when there
is no single predominant direction of vortex flux. It is tempting to ascribe
the first-order character of the phase transition to this additional
qualitative difference between the configurations in the confined and
deconfined phases (branching vs.~non-branching in space slices); on the
other hand, since the branching probability does not strictly behave as an 
order parameter, this surmise must be treated with caution. A still more 
detailed understanding of the dynamics at the deconfinement phase transition 
would be desirable. In the $SU(2)$ case, where no vortex branching can occur,
the transition is second order, at least to the level of
accuracy which was sufficient to ascertain the first order character in
the $SU(3)$ case.

As far as the confinement properties are concerned, an interesting extension
of the present model would lie in the treatment of a higher number of colours.
In particular, it would be interesting to study whether the deconfinement
phase transition becomes more strongly first order, as it does in full
lattice Yang-Mills theory \cite{highern}. Also, it has recently been argued
\cite{largen} that a description of the infrared sector of Yang-Mills
theory in terms of vortex world-surfaces of a well-defined, finite thickness
which percolate throughout space-time may become less appropriate as the number
of colours $N_c $ increases. It would be interesting to investigate whether,
and at what $N_c $, signals of this can be detected. At least in terms of the
confinement properties of $SU(3)$ Yang-Mills theory studied in the present
work, the random vortex world-surface model still seems to perform well.

A further, phenomenologically important point of study concerns the baryonic
potential, specifically whether it satisfies an area law of the $\Delta $
type or of the Y type \cite{cornbary,forc}. This is currently being
investigated. Moreover, apart from the confinement properties, also the
topological properties of the $SU(3)$ random vortex world-surface model,
which enter the axial $U_A (1)$ anomaly, need to be considered in analogy
to the $SU(2)$ case \cite{R4}. Finally, to arrive at a comprehensive
description of $SU(3)$ infrared strong interaction physics, the coupling
to quark fields and the concomitant generation of a chiral condensate
must be investigated (cf.~ref.~\cite{R5} for the $SU(2)$ model).


\section*{Acknowledgments}
\label{acknowledge}
The authors gratefully acknowledge science+computing ag, T\"ubingen, for
providing computational resources.


\end{document}